\documentclass[12pt]{article}
\usepackage{url}
\usepackage{amsmath,bm}%
\usepackage{amsfonts}%
\usepackage{geometry}
\usepackage{setspace}
\newcommand{\beaa}{\begin{eqnarray*}}
\newcommand{\eeaa}{\end{eqnarray*}}
\newcommand{\bea}{\begin{eqnarray}}
\newcommand{\eea}{\end{eqnarray}}

\usepackage{amsmath}
\usepackage{hyperref}
\usepackage{array} 
\usepackage{bm}
\newcommand\independent{\protect\mathpalette{\protect\independenT}{\perp}}
\def\independenT#1#2{\mathrel{\rlap{$#1#2$}\mkern2mu{#1#2}}} 
\usepackage{tabularx}
\usepackage{multirow}
\usepackage{dcolumn}
\numberwithin{equation}{section}
\renewcommand{\theequation}{\thesection.\arabic{equation}}

\newtheorem{thm}{Theorem}[section]
\renewcommand{\thethm}{\arabic{section}.\arabic{thm}}

\newtheorem{lemma}[thm]{Lemma}

\newtheorem{rmk}[thm]{Remark}
\newtheorem{assu}[thm]{Assumption}

\usepackage{graphicx}
\usepackage{float}
\usepackage{rotating}
\begin{document}

\title{\bf{Transformed Central Quantile Subspace}}
\date{}
\maketitle

\begin{center}
{\Large{\textbf{Eliana Christou}}\\
\textbf{University of North Carolina at Charlotte}}
\end{center}

\noindent \textbf{Abstract:}  Quantile regression (QR) is becoming increasingly popular due to its relevance in many scientific investigations.  However, application of QR can become very challenging when dealing with high-dimensional data, making it necessary to use \textit{dimension reduction techniques}.  Existing dimension reduction techniques focus on the entire conditional distribution.  We turn our attention to dimension reduction techniques for \textit{conditional quantiles} and introduce a method that serves as an intermediate step between linear and nonlinear dimension reduction.  The idea is to apply existing linear dimension reduction techniques on the \textit{transformed} predictors.  The proposed estimator, which is shown to be $\sqrt{n}$-consistent, is demonstrated through simulation examples and real data applications.  Our results suggest that this method outperforms linear dimension reduction for conditional quantiles.            


\section{Introduction} \label{Intro}

Quantile regression (QR) was first introduced by Koenker and Bassett (1978) and since then, it has received a lot of attention.  Letting $Q_{\tau}(Y|\mathbf{x}) \equiv Q_{\tau}(Y|\mathbf{X}=\mathbf{x})= \inf \{y: \Pr(Y \leq y|\mathbf{X}=\mathbf{x}) \geq \tau\}$ denote the $\tau$-th conditional quantile of the response $Y$ given a $p$-dimensional vector of predictors $\mathbf{X}=\mathbf{x}$, Koenker and Bassett (1978) considered the linear QR model $Q_{\tau}(Y|\mathbf{x})=\alpha_{\tau}+\boldsymbol{\beta}_{\tau}^{\top}\mathbf{x}$, where $\alpha_{\tau} \in \mathbb{R}$ and $\boldsymbol{\beta}_{\tau} \in \mathbb{R}^{p}$.  Specifically, they used the representation 
\begin{eqnarray*}
Q_{\tau}(Y|\mathbf{x})= \arg \min_{q} E\{\rho_{\tau}(Y-q)|\mathbf{X}=\mathbf{x}\},
\end{eqnarray*}
where, for $0<\tau<1$, the loss function $\rho_{\tau}(\cdot)$ is defined as $\rho_{\tau}(u)=\{\tau - I(u<0)\}u$, to define the estimator $(\widehat{\alpha}_{\tau}, \widehat{\boldsymbol{\beta}}_{\tau})$ as 
\begin{eqnarray*} \label{linear_QR}
(\widehat{\alpha}_{\tau}, \widehat{\boldsymbol{\beta}}_{\tau})= \arg \min_{(a_{\tau},\mathbf{b}_{\tau})} \sum_{i=1}^{n} \rho_{\tau}(Y_{i}-a_{\tau}-\mathbf{b}_{\tau}^{\top}\mathbf{X}_{i}),
\end{eqnarray*}
for $\{Y_{i},\mathbf{X}_{i}\}_{i=1}^{n}$ independent and identically distributed (i.i.d.) observations.  

Because the linearity assumption is quite strict, several authors considered the completely flexible nonparametric QR model; see, for example, Chaudhuri (1991), Yu and Jones (1998), Kong et al. (2010), and Guerre and Sabbah (2012).  However, when the number of the predictors is large, \textit{dimension reduction} techniques are used for low-dimensional smoothing without specifying any parametric or nonparametric regression relation.  

Existing literature considers linear dimension reduction techniques for the conditional quantile of the response given the predictors, that is, techniques that find the fewest \textit{linear combinations} of $\mathbf{X}$ that contain all the information on that function.  Wu et al. (2010), Kong and Xia (2012), and Christou and Akritas (2016, 2018) considered the single index quantile regression (SIQR) model and proposed a method for estimating the vector of the coefficients of the linear combination of $\mathbf{X}$.  Kong and Xia (2014) proposed an adaptive composite QR approach, which can be used for estimating multiple linear combinations of $\mathbf{X}$ that contain all the information about the conditional quantile, while Luo et al. (2014) introduced a sufficient dimension reduction method that targets any statistical functional of interest, including the conditional quantile.  Christou (2019) introduced the concept of the $\tau$th central quantile subspace ($\tau$-CQS) and proposed an algorithm for estimating it, which has good finite sample performance, is computationally inexpensive, and can be easily extended to any statistical functional of interest.    

All of the above methods are designed for extracting linear subspaces for conditional quantiles, and therefore, are unable to find important nonlinear features.  In this paper, we propose the \textit{first work} about \textit{transformed dimension reduction} for conditional quantiles.  Specifically, we propose an intermediate step between linear and nonlinear dimension reduction for conditional quantiles, which is based on Wang et al. (2014)'s methodology.  The idea is to transform the predictors monotonically and then use existing \textit{linear} dimension reduction techniques on the transformed variables.  In Section \ref{sec2} we review the $\tau$-CQS, while in Section \ref{sec3} we extend it to the transformed $\tau$-CQS.  Section \ref{Sim} presents results from several simulation examples and real data applications, and Section \ref{Disc} concludes.       

\section{The $\tau$th Central Quantile Subspace}\label{sec2}

We start by recalling the $\tau$-CQS from Christou (2019).    Assume that $Y \independent Q_{\tau}(Y|\mathbf{X}) | \mathbf{B}_{\tau}^{\top}\mathbf{X}$, where $\mathbf{B}_{\tau}$ denotes a $p \times d_{\tau,1}$ matrix, $d_{\tau,1} \leq p$.  The space spanned by $\mathbf{B}_{\tau}$, denoted by $\mathcal{S}(\mathbf{B}_{\tau})$, is a \textit{$\tau$th quantile dimension reduction subspace} for the regression of $Y$ on $\mathbf{X}$.  The intersection of all $\tau$th quantile dimension reduction subspaces is called the \textit{$\tau$th central quantile subspace} ($\tau$-CQS), denoted by $\mathcal{S}_{Q_{\tau}(Y|\mathbf{X})}$, and with dimension $d_{Q_{\tau}(Y|\mathbf{X})}$.      

The following notation will be used.  The central subspace (CS)\footnote{A dimension reduction subspace is the column space of any $p \times d_{1}$ matrix $\mathbf{A}$, $d_{1} \leq p$, such that $Y$ and $\mathbf{X}$ are conditionally independent given $\mathbf{A}^{\top}\mathbf{X}$, and the central subspace (CS) is the dimension reduction subspace with the smallest dimension (Li 1991).} is spanned by the $p \times d_{1}$ matrix $\mathbf{A}$, i.e., $\mathcal{S}_{Y|\mathbf{X}}=\mathcal{S}(\mathbf{A})$, and, for a given $\tau$, the $\tau$-CQS is spanned by the $p \times d_{\tau,1}$ matrix $\mathbf{B}_{\tau}$, i.e., $\mathcal{S}_{Q_{\tau}(Y|\mathbf{X})}=\mathcal{S}(\mathbf{B}_{\tau})$.  

The goal is to describe the dependence on $\mathbf{X}$ of the $\tau$th conditional quantile of $Y$ given $\mathbf{X}$.  Therefore, Christou (2019) proposed the following to determine the matrix of coefficients for the linear combination $\mathbf{B}_{\tau}^{\top}\mathbf{X}$.
\begin{assu} \label{Ass1}
For a given $\tau$, the conditional expectation $E(\mathbf{b}_{\tau}^{\top}\mathbf{X}|\mathbf{B}_{\tau}^{\top}\mathbf{X})$ is linear in $\mathbf{B}_{\tau}^{\top}\mathbf{X}$ for every $\mathbf{b}_{\tau} \in \mathbb{R}^{p}$. 
\end{assu}
\begin{enumerate}
\item[(a)] Under Assumption \ref{Ass1} and if $d_{Q_{\tau}(Y|\mathbf{X})}=1$, i.e., $\mathbf{B}_{\tau}$ is a $p \times 1$ vector, then $\boldsymbol{\beta}_{\tau}^* \in \mathcal{S}_{Q_{\tau}(Y|\mathbf{X})}$ for 
\begin{eqnarray}\label{OLS}
(\alpha_{\tau}^*, \boldsymbol{\beta}^*_{\tau})= \arg \min_{(a_{\tau},\mathbf{b}_{\tau})} E\{Q_{\tau}(Y|\mathbf{A}^{\top}\mathbf{X})-a_{\tau}-\mathbf{b}_{\tau}^{\top}\mathbf{X}\}^2,
\end{eqnarray}
and $\mathbf{A}$ is such that $\mathcal{S}(\mathbf{A})=\mathcal{S}_{Y|\mathbf{X}}$.

\item[(b)] If $d_{Q_{\tau}(Y|\mathbf{X})}>1$, then the vector $\boldsymbol{\beta}_{\tau}^*$, defined in (\ref{OLS}), is inconsistent for estimating $\mathcal{S}_{Q_{\tau}(Y|\mathbf{X})}$, and a different approach is necessary to produce more vectors in $\mathcal{S}_{Q_{\tau}(Y|\mathbf{X})}$.  Under Assumption \ref{Ass1}, and the assumption that $U_{\tau}$ is a measurable function of $\mathbf{B}_{\tau}^{\top}\mathbf{X}$, $E\{Q_{\tau}(Y|U_{\tau})\mathbf{X}\} \in  \mathcal{S}_{Q_{\tau}(Y|\mathbf{X})}$, provided that $Q_{\tau}(Y|U_{\tau})\mathbf{X}$ is integrable.
This suggests that if we know one vector $\boldsymbol{\beta}_{\tau,0} \in  \mathcal{S}_{Q_{\tau}(Y|\mathbf{X})}$, then we can find more vectors in $\mathcal{S}_{Q_{\tau}(Y|\mathbf{X})}$ using
\begin{eqnarray*}
\boldsymbol{\beta}_{\tau,j}=E[Q_{\tau}\{Y|u_{\tau}(\boldsymbol{\beta}_{\tau,j-1}^{\top}\mathbf{X})\}\mathbf{X}] \in  \mathcal{S}_{Q_{\tau}(Y|\mathbf{X})},
\end{eqnarray*}
for $j=1, 2, \dots$.  Christou (2019) used $\boldsymbol{\beta}_{\tau,0}=\boldsymbol{\beta}_{\tau}^*$, defined in (\ref{OLS}), and $u_{\tau}(t)=t$.  This suggests the following procedure: Set $\boldsymbol{\beta}_{\tau,0}=\boldsymbol{\beta}_{\tau}^*$ and, for $j=1, 2, \dots, p-1$, let $\boldsymbol{\beta}_{\tau,j}=E\{Q_{\tau}(Y|\boldsymbol{\beta}_{\tau,j-1}^{\top}\mathbf{X})\mathbf{X}\}$.  Then, $\boldsymbol{\beta}_{\tau,j} \in \mathcal{S}_{Q_{\tau}(Y|\mathbf{X})}$, $j=0,1,\dots, p-1$, and $d_{Q_{\tau}(Y|\mathbf{X})}$ vectors of the sequence $\boldsymbol{\beta}_{\tau,0}, \dots, \boldsymbol{\beta}_{\tau, p-1}$ are linearly independent and form the $\tau$-CQS.  However, to obtain linearly independent vectors, let $\mathbf{V}_{\tau}$ be the $p \times p$ matrix with column vectors $\boldsymbol{\beta}_{\tau,0}, \dots, \boldsymbol{\beta}_{\tau, p-1}$, and perform an eigenvalue decomposition on $\mathbf{V}_{\tau}\mathbf{V}_{\tau}^{\top}$ to select the $d_{Q_{\tau}(Y|\mathbf{X})}$ linearly independent eigenvectors $\mathbf{v}_{\tau,1}, \dots, \mathbf{v}_{\tau,d_{Q_{\tau}(Y|\mathbf{X})}}$ corresponding to the $d_{Q_{\tau}(Y|\mathbf{X})}$ non-zero eigenvalues.  Then, $(\mathbf{v}_{\tau,1}, \dots, \mathbf{v}_{\tau, d_{Q_{\tau}(Y|\mathbf{X})}}) \in \mathcal{S}_{Q_{\tau}(Y|\mathbf{X})}$.   
\end{enumerate}

The above procedure, which is explained in greater detail in Christou (2019), leads to the following estimation method.  First, use a standard dimension reduction technique to estimate $\mathbf{A}$ by $\widehat{\mathbf{A}}$ and form the new $d_{1} \times 1$ predictor vector $\widehat{\mathbf{A}}^{\top}\mathbf{X}$.  Next, use the data $\{Y_{i}, \mathbf{X}_{i}\}_{i=1}^{n}$ to estimate $\boldsymbol{\beta}_{\tau}^*$ by   
\begin{eqnarray*} \label{OLS_sample}
(\widehat{a}_{\tau},\widehat{\boldsymbol{\beta}}_{\tau})=\arg \min_{(a_{\tau},\mathbf{b}_{\tau})}\sum_{i=1}^{n}\{\widehat{Q}_{\tau}(Y|\widehat{\mathbf{A}}^{\top}\mathbf{X}_{i})-a_{\tau}-\mathbf{b}_{\tau}^{\top}\mathbf{X}_{i}\}^2,
\end{eqnarray*}
where $\widehat{Q}_{\tau}(Y|\widehat{\mathbf{A}}^{\top}\mathbf{X}_{i})$ is a nonparametric estimate of $Q_{\tau}(Y|\widehat{\mathbf{A}}^{\top}\mathbf{X}_{i})$.  Note that $\widehat{\boldsymbol{\beta}}_{\tau}$ is the slope estimate in an ordinary least squares regression of $\widehat{Q}_{\tau}(Y|\widehat{\mathbf{A}}^{\top}\mathbf{X})$ on $\mathbf{X}$.  There are many ways to estimate $Q_{\tau}(Y|\widehat{\mathbf{A}}^{\top}\mathbf{X}_{i})$; Christou (2019) used the local linear conditional quantile estimation method introduced in Guerre and Sabbah (2012).  Specifically, take $\widehat{Q}_{\tau}(Y|\widehat{\mathbf{A}}^{\top}\mathbf{X}_{i})=\widehat{q}_{\tau}(\mathbf{X}_{i})$, where 
\begin{eqnarray} \label{llqr}
(\widehat{q}_{\tau}(\mathbf{X}_{i}), \widehat{\mathbf{s}}_{\tau}(\mathbf{X}_{i})) = \arg \min_{(q_{\tau}, \mathbf{s}_{\tau})} \sum_{k=1}^{n} \rho_{\tau} \{Y_{k}-q_{\tau}-\mathbf{s}_{\tau}^{\top}\widehat{\mathbf{A}}^{\top}(\mathbf{X}_{k}-\mathbf{X}_{i})\} K \left\{\frac{\widehat{\mathbf{A}}^{\top}(\mathbf{X}_{k}-\mathbf{X}_{i})}{h}\right\},
\end{eqnarray}
for $K(\cdot)$ a $d_{1}$-dimensional kernel function and $h>0$ a bandwidth. 

Following, if $d_{Q_{\tau}(Y|\mathbf{X})}=1$, then stop and report $\widehat{\boldsymbol{\beta}}_{\tau}$ as the estimated basis vector for $\mathcal{S}_{Q_{\tau}(Y|\mathbf{X})}$.  If $d_{Q_{\tau}(Y|\mathbf{X})}>1$, then set $\widehat{\boldsymbol{\beta}}_{\tau,0}=\widehat{\boldsymbol{\beta}}_{\tau}$ and form the vectors $\widehat{\boldsymbol{\beta}}_{\tau,j}=n^{-1}\sum_{i=1}^{n}\widehat{Q}_{\tau}(Y|\widehat{\boldsymbol{\beta}}_{\tau,j-1}^{\top}\mathbf{X}_{i})\mathbf{X}_{i}$, for $j=1, \dots, p-1$, where $\widehat{Q}_{\tau}(Y|\widehat{\boldsymbol{\beta}}_{\tau,j-1}^{\top}\mathbf{X}_{i})$ is the local linear conditional quantile estimate of $Q_{\tau}(Y|\widehat{\boldsymbol{\beta}}_{\tau,j-1}^{\top}\mathbf{X}_{i})$, i.e., $\widehat{Q}_{\tau}(Y|\widehat{\boldsymbol{\beta}}_{\tau,j-1}^{\top}\mathbf{X}_{i})=\widehat{q}_{\tau}(\mathbf{X}_{i})$ from (\ref{llqr}) but $\widehat{\mathbf{A}}$ is replaced with $\widehat{\boldsymbol{\beta}}_{\tau, j-1}$.  Finally, form the $p \times p$ matrix $\widehat{\mathbf{V}}_{\tau}=(\widehat{\boldsymbol{\beta}}_{\tau,0}, \dots, \widehat{\boldsymbol{\beta}}_{\tau,p-1})$ and choose the eigenvectors $\widehat{\mathbf{v}}_{\tau,k}$, $k=1, \dots, d_{Q_{\tau}(Y|\mathbf{X})}$, corresponding to the $d_{Q_{\tau}(Y|\mathbf{X})}$ largest eigenvalues of $\widehat{\mathbf{V}}_{\tau}\widehat{\mathbf{V}}_{\tau}^{\top}$.  Below is the algorithm.      

\hrulefill

\noindent \textbf{Algorithm 1:} Let $\{Y_{i}, \mathbf{X}_{i}\}_{i=1}^{n}$ i.i.d. observations and fix $\tau \in (0,1)$.    
\begin{enumerate}
\item Use sliced inverse regression (SIR) of Li (1991) or a similar dimension reduction technique to estimate the $p \times d_{1}$ basis matrix $\mathbf{A}$ of the CS, denoted by $\widehat{\mathbf{A}}$, and form the new sufficient predictors $\widehat{\mathbf{A}}^{\top}\mathbf{X}_{i}$, $i=1, \dots, n$.
\item For each $i=1, \dots, n$, use the local linear conditional quantile estimation method of Guerre and Sabbah (2012) to estimate $Q_{\tau}(Y|\widehat{\mathbf{A}}^{\top}\mathbf{X}_{i})$.  Specifically, take $\widehat{Q}_{\tau}(Y|\widehat{\mathbf{A}}^{\top}\mathbf{X}_{i})=\widehat{q}_{\tau}(\mathbf{X}_{i})$, where $\widehat{q}_{\tau}(\mathbf{X}_{i})$ is given in (\ref{llqr}).    
\item Take $\widehat{\boldsymbol{\beta}}_{\tau}$ to be 
\begin{eqnarray*} 
(\widehat{a}_{\tau},\widehat{\boldsymbol{\beta}}_{\tau})=\arg \min_{(a_{\tau},\mathbf{b}_{\tau})}\sum_{i=1}^{n}\{\widehat{Q}_{\tau}(Y|\widehat{\mathbf{A}}^{\top}\mathbf{X}_{i})-a_{\tau}-\mathbf{b}_{\tau}^{\top}\mathbf{X}_{i}\}^2.
\end{eqnarray*}
\item If $d_{Q_{\tau}(Y|\mathbf{X})}=1$, stop and report $\widehat{\boldsymbol{\beta}}_{\tau}$ as the estimated basis vector for $\mathcal{S}_{Q_{\tau}(Y|\mathbf{X})}$.  Otherwise, move to Step 5.
\item Set $\widehat{\boldsymbol{\beta}}_{\tau,0}=\widehat{\boldsymbol{\beta}}_{\tau}$.
\item Given $j$, for $j=1, \dots, p-1$,
\begin{enumerate}
\item form the predictors $\widehat{\boldsymbol{\beta}}_{\tau, j-1}^{\top}\mathbf{X}_{i}$, $i=1, \dots, n$, and use the local linear conditional quantile estimation method of Guerre and Sabah (2012) to estimate $Q_{\tau}(Y|\widehat{\boldsymbol{\beta}}_{\tau, j-1}^{\top}\mathbf{X}_{i})$.  Specifically, take $\widehat{Q}_{\tau}(Y|\widehat{\boldsymbol{\beta}}_{\tau, j-1}^{\top}\mathbf{X}_{i})=\widehat{q}_{\tau}(\mathbf{X}_{i})$, where $\widehat{q}_{\tau}(\mathbf{X}_{i})$ is given in (\ref{llqr}), except that we replace $\widehat{\mathbf{A}}$ by $\widehat{\boldsymbol{\beta}}_{\tau, j-1}$.  This leads to a univariate kernel function $K(\cdot)$.   
\item let $\widehat{\boldsymbol{\beta}}_{\tau, j}=n^{-1} \sum_{i=1}^{n} \widehat{Q}_{\tau}(Y|\widehat{\boldsymbol{\beta}}_{\tau, j-1}^{\top}\mathbf{X}_{i})\mathbf{X}_{i}$.
\end{enumerate}
\item Repeat Step 6 for $j=1, \dots, p-1$. 
\item Let $\widehat{\mathbf{V}}_{\tau}$ be the $p \times p$ matrix with column vectors $\widehat{\boldsymbol{\beta}}_{\tau, j}$, $j=0, 1, \dots, p-1$, that is, $\widehat{\mathbf{V}}_{\tau}=(\widehat{\boldsymbol{\beta}}_{\tau,0}, \dots, \widehat{\boldsymbol{\beta}}_{\tau, p-1})$, and choose the eigenvectors $\widehat{\mathbf{v}}_{\tau,k}$, $k=1, \dots, d_{Q_{\tau}(Y|\mathbf{X})}$, corresponding to the $d_{Q_{\tau}(Y|\mathbf{X})}$ largest eigenvalues of $\widehat{\mathbf{V}}_{\tau}\widehat{\mathbf{V}}_{\tau}^{\top}$.  Then, $\widehat{\mathbf{B}}_{\tau}=(\widehat{\mathbf{v}}_{\tau,1}, \dots, \widehat{\mathbf{v}}_{\tau, d_{Q_{\tau}(Y|\mathbf{X})}})$ is an estimated basis matrix for $\mathcal{S}_{Q_{\tau}(Y|\mathbf{X})}$.   
\end{enumerate}
\hrulefill

Note that, Step 1 of the above algorithm performs an initial dimension reduction and gives the same $\widehat{\mathbf{A}}$ for all choices of $\tau$.  This is then converted into an estimate of $\mathbf{B}_{\tau}$, which now depends on a given value of $\tau$.  

\section{The Transformed $\tau$th Central Quantile Subspace} \label{sec3}

Section \ref{sec2} focuses on methods for extracting \textit{linear} subspaces for conditional quantiles, and therefore, are unable to find important nonlinear features.  That is the reason we turn our attention to nonlinear dimension reduction methods that seek an arbitrary function $\psi_{\tau}$ from $\mathbb{R}^{p}$ to $\mathbb{R}^{d_{\tau}}$ such that 
\begin{eqnarray} \label{nonlinear}
Y \independent Q_{\tau}(Y|\mathbf{X}) | \psi_{\tau}(\mathbf{X}).
\end{eqnarray}
Nonlinear dimension reduction techniques can potentially achieve greater dimension reduction.  This is because ``if the data are concentrated on a nonlinear low-dimensional space, the linear dimension-reduction subspace to be estimated is often of a very large dimension" (Wang et al. 2014, p. 816).  To the best of our knowledge, there is no work about nonlinear dimension reduction for the conditional quantiles.

To better understand the difference between linear and nonlinear dimension reduction for conditional quantiles, we borrow an example from Wang et al. (2014).  Let $Y=X_{1}+X_{2}^2+X_{3}^3+\exp(X_{4})+\varepsilon$, where $p=6$ and all the predictors and $\varepsilon$ are independent. Linear dimension reduction on the conditional quantiles yields $d_{Q_{\tau}(Y|\mathbf{X})}=4$, for every $\tau \in (0,1)$, whereas nonlinear dimension reduction on the conditional quantiles gives $\psi_{\tau}(\mathbf{X})=X_{1}+X_{2}^2+X_{3}^3+\exp(X_{4})$ and yields a one-dimensional nonlinear $\tau$-CQS.  However, the transition from linear to nonlinear dimension reduction for the conditional quantiles results in loss of interpretability.  Furthermore, linear dimension reduction techniques are usually used as an initial dimension reduction before applying another, more sophisticated method, while nonlinear dimension reduction solves the entire problem in one step.  A good compromise between the two is offered by the \textit{transformed dimension reduction}, introduced by Wang et al. (2014).  

Let $p$ monotone univariate functions $g_{1}, \dots, g_{p}$ and write $\mathbf{g}=(g_{1}, \dots, g_{p})^{\top}$.  Then, relation (\ref{nonlinear}) is equivalent to $Y \independent Q_{\tau}\{Y|\mathbf{g}(\mathbf{X})\} | \phi_{\tau}\{\mathbf{g}(\mathbf{X})\}$, for another function $\phi_{\tau}$ from $\mathbb{R}^{p}$ to $\mathbb{R}^{d_{\tau,2}}$.  To generalize linear dimension reduction for conditional quantiles, while preserving its simplicity, assume further that $\phi_{\tau}$ is linear, that is, there exists a $p \times d_{\tau,2}$ matrix $\mathbf{T}_{\tau}$ such that 
\begin{eqnarray} \label{transformed}
Y \independent Q_{\tau}\{Y|\mathbf{g}(\mathbf{X})\} | \mathbf{T}_{\tau}^{\top} \mathbf{g}(\mathbf{X}).
\end{eqnarray}
The space spanned by the matrix $\mathbf{T}_{\tau}$ is called the \textit{transformed $\tau$th quantile dimension reduction subspace} for the regression of $Y$ on $\mathbf{X}$ with respect to $\mathbf{g}$.  The \textit{transformed $\tau$-CQS}, denoted by $\mathcal{S}_{Q_{\tau}\{Y|\mathbf{g}(\mathbf{X})\}}$, is defined to be the intersection of all transformed $\tau$th quantile dimension reduction subspaces, with dimension denoted by $d_{Q_{\tau}\{Y|\mathbf{g}(\mathbf{X})\}}$.  

To better understand this intermediate step, consider again the example $Y=X_{1}+X_{2}^2+X_{3}^3+\exp(X_{4})+\varepsilon$, where $p=6$ and all the predictors and $\varepsilon$ are independent.  If we take $g_{j}(X_{j})=X_{j}$ for $j=1,2,5,6$, $g_{3}(X_{3})=X_{3}^3$, and $g_{4}(X_{4})=\exp(X_{4})$, then the model can be re-expressed as $Y=g_{1}(X_{1})+\{g_{2}(X_{2})\}^2+g_{3}(X_{3})+g_{4}(X_{4})+\varepsilon$.  This means that the transformed $\tau$-CQS yields two dimensions, i.e., (1,0,1,1,0,0) and (0,1,0,0,0,0), for every $\tau \in (0,1)$.  Therefore, transformed dimension reduction achieves greater dimension reduction than linear dimension reduction, while retains the flexibility of a nonlinear dimension reduction.     

Model (\ref{transformed}) suggests that we can apply linear dimension reduction techniques using $\mathbf{g}(\mathbf{X})$ instead of $\mathbf{X}$.  Specifically, we can apply Christou (2019)'s algorithm on the transformed predictors $\mathbf{g}(\mathbf{X})$.  To do this, $\mathbf{g}$ needs to be specified.  Wang et al. (2014) proposed, among other methods, to assume that the transformed vector $\mathbf{g}(\mathbf{X})=(g_{1}(X_{1}), \dots, g_{p}(X_{p}))^{\top}$ is multivariate Gaussian and each $g_{j}$ is monotonically increasing.  To ensure identifiability, assume that $\boldsymbol{\mu}_{\mathbf{g}}=E\{\mathbf{g}(\mathbf{X})\}=\mathbf{0}$ and $\mathbf{\Sigma}_{\mathbf{g}}=Cov\{\mathbf{g}(\mathbf{X})\}$ is a correlation matrix whose diagonal entries equal unity.  Under the identifiability condition, $g_{j}(\cdot)=\Phi^{-1}\{F_{j}(\cdot)\}$, where $\Phi(\cdot)$ and $F_{j}(\cdot)$ denote the standard normal distribution function and the marginal distribution function of $X_{j}$, respectively.  In the sample level, let $r_{ij}$ denote the ranks of the $n$ observations for the $j$th predictor.  Define the normal scores $u_{ij}=\Phi^{-1}\{\widehat{F}_{j}(X_{ij})\}=\Phi^{-1} \{r_{ij}/(n+1)\}$, where $\widehat{F}_{j}(\cdot)$ denotes the empirical marginal distribution function of $X_{j}$, and let $\widehat{\mathbf{g}}(\mathbf{X}_{i})=(u_{i1}, \dots, u_{ip})^{\top}$.  Then, replace the observations $\mathbf{X}_{i}$ with $\widehat{\mathbf{g}}(\mathbf{X}_{i})$, for $i=1, \dots, n$, and apply Algorithm 1 to estimate a basis matrix for $\mathcal{S}_{Q_{\tau}\{Y|\mathbf{g}(\mathbf{X})\}}$. 

The following notation will be used.  The transformed CS\footnote{A transformed dimension reduction subspace is the column space of any $p \times d_{2}$ matrix $\mathbf{\Gamma}$, $d_{2} \leq p$, such that $Y$ and $\mathbf{g}(\mathbf{X})$ are conditionally independent given $\mathbf{\Gamma}^{\top}\mathbf{g}(\mathbf{X})$, and the transformed CS is the transformed dimension reduction subspace with the smallest dimension (Wang et al. 2014)} is spanned by the $p \times d_{2}$ matrix $\mathbf{\Gamma}$, i.e., $\mathcal{S}_{Y|\mathbf{g}(\mathbf{X})}=\mathcal{S}(\mathbf{\Gamma})$, and, for a given $\tau$, the transformed $\tau$-CQS is spanned by the $p \times d_{\tau,2}$ matrix $\mathbf{T}_{\tau}$, i.e., $\mathcal{S}_{Q_{\tau}\{Y|\mathbf{g}(\mathbf{X})\}}=\mathcal{S}(\mathbf{T}_{\tau})$.    
 
 \hrulefill
 
\noindent \textbf{Algorithm 2:}  Let $\{Y_{i}, \mathbf{X}_{i}\}_{i=1}^{n}$ i.i.d. observations and fix $\tau \in (0,1)$.   
\begin{enumerate}
\item For $i=1, \dots, n$, and $j=1, \dots, p$, let $r_{ij}$ denote the ranks of the $n$ observations from the $j$th predictor.  Define the normal scores $u_{ij}=\Phi^{-1} \{r_{ij}/(n+1)\}$, where $\Phi(\cdot)$ denotes the standard normal distribution function, and denote $\widehat{\mathbf{g}}(\mathbf{X}_{i})=(u_{i1}, \dots, u_{ip})^{\top}$.
\item Apply Algorithm 1 using $\{Y_{i}, \widehat{\mathbf{g}}(\mathbf{X}_{i})\}_{i=1}^{n}$ to obtain eigenvectors $\widehat{\mathbf{w}}_{\tau, k}$, $k=1, \dots, d_{Q_{\tau}\{Y|\mathbf{g}(\mathbf{X})\}}$.  Specifically,
\begin{enumerate}
\item Use SIR of Li (1991) to estimate the $p \times d_{2}$ basis matrix $\boldsymbol{\Gamma}$ of the transformed CS, denoted by $\widehat{\boldsymbol{\Gamma}}$, and form the new sufficient predictors $\widehat{\boldsymbol{\Gamma}}^{\top} \widehat{\mathbf{g}}(\mathbf{X}_{i})$, $i=1, \dots, n$.  

\item For each $i=1, \dots, n$, use the local linear conditional quantile estimation method of Guerre and Sabbah (2012) to estimate $Q_{\tau}\{Y|\widehat{\boldsymbol{\Gamma}}^{\top} \widehat{\mathbf{g}}(\mathbf{X}_{i})\}$.  Specifically, take $\widehat{Q}_{\tau}\{Y|\widehat{\boldsymbol{\Gamma}}^{\top} \widehat{\mathbf{g}}(\mathbf{X}_{i})\}=\widehat{q}_{\tau}\{\widehat{\mathbf{g}}(\mathbf{X}_{i})\}$, where $\widehat{q}_{\tau}\{\widehat{\mathbf{g}}(\mathbf{X}_{i})\}$ is given in (\ref{llqr}), except that we replace $\widehat{\mathbf{A}}$ with $\widehat{\boldsymbol{\Gamma}}$ and $\mathbf{X}_{i}$ with $\widehat{\mathbf{g}}(\mathbf{X}_{i})$.  This leads to a $d_{2}$-dimensional kernel function $K(\cdot)$.      
    
\item  Take $\widehat{\boldsymbol{\eta}}_{\tau}$ to be
\begin{eqnarray} \label{OLS_transformed}
(\widehat{\gamma}_{\tau}, \widehat{\boldsymbol{\eta}}_{\tau})= \arg \min_{(\gamma_{\tau}, \boldsymbol{\eta}_{\tau})} \sum_{i=1}^{n} [\widehat{Q}_{\tau}\{Y|\widehat{\mathbf{\Gamma}}^{\top}\widehat{\mathbf{g}}(\mathbf{X}_{i})\}-\gamma_{\tau}-\boldsymbol{\eta}_{\tau}^{\top}\widehat{\mathbf{g}}(\mathbf{X}_{i})]^2.
\end{eqnarray}

\item If $d_{Q_{\tau}\{Y|\mathbf{g}(\mathbf{X})\}}=1$, stop and report $\widehat{\boldsymbol{\eta}}_{\tau}$ as the estimated basis vector for $\mathcal{S}_{Q_{\tau}\{Y|\mathbf{g}(\mathbf{X})\}}$.  Otherwise, move to Step 2(e).

\item Set $\widehat{\boldsymbol{\eta}}_{\tau,0}=\widehat{\boldsymbol{\eta}}_{\tau}$.

\item Given $j$, for $j=1, \dots, p-1$,
\begin{enumerate}
\item form the predictors $\widehat{\boldsymbol{\eta}}^{\top}_{\tau,j-1} \widehat{\mathbf{g}}(\mathbf{X}_{i})$, $i=1, \dots, n$, and use the local linear conditional quantile estimation method of Guerre and Sabbah (2012) to estimate $Q_{\tau}\{Y|\widehat{\boldsymbol{\eta}}^{\top}_{\tau, j-1} \widehat{\mathbf{g}}(\mathbf{X}_{i})\}$.  Specifically, take $\widehat{Q}_{\tau}\{Y|\widehat{\boldsymbol{\eta}}^{\top}_{\tau, j-1} \widehat{\mathbf{g}}(\mathbf{X}_{i})\}=\widehat{q}_{\tau}\{\widehat{\mathbf{g}}(\mathbf{X}_{i})\}$, where $\widehat{q}_{\tau}\{\widehat{\mathbf{g}}(\mathbf{X}_{i})\}$ is given in (\ref{llqr}), except that we replace $\widehat{\mathbf{A}}$ by $\widehat{\boldsymbol{\eta}}_{\tau, j-1}$ and $\mathbf{X}_{i}$ by $\widehat{\mathbf{g}}(\mathbf{X}_{i})$.  This leads to a univariate kernel function $K(\cdot)$.  

\item let $\widehat{\boldsymbol{\eta}}_{\tau ,j}=n^{-1} \sum_{i=1}^{n} \widehat{Q}_{\tau}\{Y|\widehat{\boldsymbol{\eta}}^{\top}_{\tau, j-1} \widehat{\mathbf{g}}(\mathbf{X}_{i})\} \widehat{\mathbf{g}}(\mathbf{X}_{i})$. 
\end{enumerate} 

\item Let $\widehat{\mathbf{W}}_{\tau}$  be the $p \times p$ matrix with column vectors $\widehat{\boldsymbol{\eta}}_{\tau, j}$, $j=0, 1, \dots, p-1$, that is, $\widehat{\mathbf{W}}_{\tau}=(\widehat{\boldsymbol{\eta}}_{\tau,0}, \dots, \widehat{\boldsymbol{\eta}}_{\tau, p-1})$, and choose the eigenvectors $\widehat{\mathbf{w}}_{\tau, k}$, $k=1, \dots, d_{Q_{\tau}\{Y|\mathbf{g}(\mathbf{X})\}}$, corresponding to the $d_{Q_{\tau}\{Y|\mathbf{g}(\mathbf{X})\}}$ largest eigenvalues of $\widehat{\mathbf{W}}_{\tau} \widehat{\mathbf{W}}_{\tau}^{\top}$.  Then, 
\begin{eqnarray} \label{Bhat}
\widehat{\mathbf{T}}_{\tau}=(\widehat{\mathbf{w}}_{\tau, 1},  \dots, \widehat{\mathbf{w}}_{\tau, d_{Q_{\tau}\{Y|\mathbf{g}(\mathbf{X})\}}} )
\end{eqnarray}
is an estimated basis matrix for $\mathcal{S}_{Q_{\tau}\{Y|\mathbf{g}(\mathbf{X})\}}$.  
\end{enumerate}
\end{enumerate}
\hrulefill

\begin{rmk}
In practice we standardize $\widehat{\mathbf{g}}(\mathbf{X}_{i})$ using $\widehat{\mathbf{Z}}_{i}^{\mathbf{g}}=\widehat{\mathbf{\Sigma}}^{-1/2}_{\mathbf{g}}\widehat{\mathbf{g}}(\mathbf{X}_{i})$, where $\widehat{\mathbf{\Sigma}}_{\mathbf{g}}$ is the sample covariance matrix of $\widehat{\mathbf{g}}(\mathbf{X}_{i})$.
\end{rmk}

\begin{rmk}
The linearity condition of the $\tau$-CQS, i.e., Assumption \ref{Ass1}, is no longer necessary, since we assume that the transformed vector $\mathbf{g}(\mathbf{X})$ is multivariate Gaussian.
\end{rmk}

\begin{thm} \label{thm1}
For a given $\tau \in (0,1)$, assume that $Y \independent Q_{\tau}\{Y|\mathbf{g}(\mathbf{X})\} | \mathbf{T}_{\tau}^{\top}\mathbf{g}(\mathbf{X})$, where $\mathbf{T}_{\tau}$ is a $p \times d_{\tau,2}$ matrix and $d_{\tau,2} \leq p$.  If $\mathbf{g}(\mathbf{X}) \sim N(\mathbf{0}, \mathbf{\Sigma}_{\mathbf{g}})$ and Assumptions A1-A5, given in Appendix A, hold, then the column vectors of $\widehat{\mathbf{T}}_{\tau}$ are $\sqrt{n}$-consistent estimates of the directions of $\mathcal{S}_{Q_{\tau}\{Y|\mathbf{g}(\mathbf{X})\}}$, where $\widehat{\mathbf{T}}_{\tau}$ is defined in (\ref{Bhat}). 
\end{thm}

\noindent \textbf{Proof:} See Appendix \ref{AppendixB2}.  

\section{Numerical Studies}\label{Sim}

\subsection{Computational Remarks} \label{Sima}

The estimation of the basis matrices $\mathbf{A}$ and $\mathbf{\Gamma}$ of the CS and transformed CS are performed with SIR using $(Y,\mathbf{X})$ and $(Y,\widehat{\mathbf{g}}(\mathbf{X}))$, respectively, where the number of slices is chosen to be max$(10, 2p/n)$.  For the computation of the local linear conditional quantile estimator, given in (\ref{llqr}), we use the function \texttt{lprq} in the \texttt{R} package \texttt{quantreg}.  We use a Gaussian kernel and choose the bandwidth as the rule-of-thumb bandwidth given in Wu et al. (2010).  Specifically, let $h=h_{m}[\tau(1-\tau)/\phi\{\Phi^{-1}(\tau)\}^2]^{1/5}$, where $\phi(\cdot)$ and $\Phi(\cdot)$ denote the probability density and cumulative distribution functions of the standard normal distribution, respectively, and $h_{m}$ denotes the optimal bandwidth used in mean regression local estimation.  To estimate $h_{m}$ we use the function \texttt{dpill} of the \texttt{KernSmooth} package in $\texttt{R}$.     

For the estimation accuracy we use the distance measure (DM) suggested by Li et al. (2005).  Specifically, for two subspaces $\widehat{\mathbf{T}}_{\tau}$ and $\mathbf{T}_{\tau}$, we define
\begin{eqnarray*}
\text{dist}(\widehat{\mathbf{T}}_{\tau}, \mathbf{T}_{\tau}) = \left\| \widehat{\mathbf{T}}_{\tau}(\widehat{\mathbf{T}}_{\tau}^{\top} \widehat{\mathbf{T}}_{\tau})^{-1} \widehat{\mathbf{T}}_{\tau}^{\top} - \mathbf{T}_{\tau}(\mathbf{T}_{\tau}^{\top} \mathbf{T}_{\tau})^{-1} \mathbf{T}_{\tau}^{\top} \right\|,
\end{eqnarray*}
where $\left\| \cdot \right\|$ is the Euclidean norm, that is, the maximum singular value of a matrix.  Smaller values of the DM indicate better estimation accuracy.  We also report the trace correlation coefficient (TCC), defined as $(d_{\tau,2}^{-1} \sum_{i=1}^{d_{\tau,2}} \lambda_{i}^2)^{1/2}$, where $1 \geq \lambda_{1}^2 \geq \cdots \geq \lambda_{d_{\tau,2}}^{2} \geq 0$ are the eigenvalues of the matrix $\widehat{\mathbf{T}}_{0,\tau}^{\top} \mathbf{T}_{0,\tau} \mathbf{T}_{0,\tau}^{\top} \widehat{\mathbf{T}}_{0,\tau}$, with $\widehat{\mathbf{T}}_{0,\tau}$ and $\mathbf{T}_{0,\tau}$ denote the orthonormalized versions of $\widehat{\mathbf{T}}_{\tau}$ and $\mathbf{T}_{\tau}$, respectively.  Since this is a correlation measure, a value closer to one indicates better estimation accuracy of the subspace spanned by the matrix $\mathbf{T}_{\tau}$.       

All simulation results are based on $N=100$ iterations.  Unless otherwise stated, the sample size is chosen to be $n=600$, and the quantiles under consideration are $\tau=0.1, 0.25, 0.5, 0.75$, and 0.9.    

\textbf{Note:} The purpose of this paper is to illustrate the advantage of the transformed $\tau$-CQS over the linear $\tau$-CQS.  Therefore, for the following simulation examples, we will compare the proposed methodology with that of Christou (2019).  The comparison between the $\tau$-CQS and other existing linear dimension reduction techniques for conditional quantiles, such as that of Kong and Xia (2014) and Luo et al. (2014), was already performed in Christou (2019).  

\subsection{Simulation Results} \label{Simb}

\noindent \textbf{Example 1:} We begin by considering the overall performance of the proposed transformed $\tau$-CQS for different choices of $n$ and $p$.  The data is generated according to the following model
\begin{eqnarray*}
Y=2\exp\left(\frac{X_{1}}{3}\right)+\frac{X_{2}^3}{3}+X_{3}+X_{4}+0.5\varepsilon,
\end{eqnarray*}
where $\mathbf{X}=(X_{1}, \dots, X_{p})^{\top}$ and the error $\varepsilon$ are generated according to a standard normal distribution.  The sample size is given by $n=400, 600$ or 800, and the number of predictors is $p=10, 20$ or 40.  For $g_{1}(X_{1})=2\exp(X_{1}/3)$, $g_{2}(X_{2})=X_{2}^3/3$, $g_{3}(X_{3})=X_{3}$, and $g_{4}(X_{4})=X_{4}$, the transformed $\tau$-CQS is spanned by $(1,1,1,1,0, \dots, 0)^{\top}$, for $\tau=0.1, 0.25, 0.5, 0.75, 0.9$.  Table \ref{tab:example1} presents the mean and standard deviation of DM and TCC for the estimation of the transformed $\tau$-CQS.  As expected, the estimation accuracy increases with $n$ and decreases with $p$.        

\setlength{\arrayrulewidth}{.15em}
\begin{table}[h!]
\caption{\label{tab:example1}\it{Mean (and standard deviation) of the estimation accuracy for $\widehat{\mathbf{T}}_{\tau}$, $\tau=0.1, 0.25, 0.5, 0.75, 0.9$, for Example 1.}}
\begin{center}
\begin{tabularx}{\textwidth}{cc|@{\extracolsep\fill}ccccc}
\hline
$n$ & $p$ & 0.1 & 0.25 & 0.5 & 0.75 & 0.9 \\
\hline
DM \\
\hline
400 & 10 & 0.217 (0.066) & 0.195 (0.051) & 0.187 (0.042)	& 0.192 (0.044)	 & 0.209 (0.059) \\
& 20 & 0.318 (0.076)	& 0.278 (0.057)	& 0.257 (0.046)	& 0.271 (0.061)	 & 0.309 (0.081) \\
& 40 & 0.450 (0.106) & 0.397 (0.095) & 0.377 (0.089) & 0.403 (0.092) & 0.460 (0.102)  \\
\hline
600 & 10 & 0.172 (0.040)	& 0.164 (0.035)	& 0.165 (0.033)	& 0.169 (0.041)	 & 0.176 (0.041) \\
& 20 & 0.258 (0.078) & 0.212 (0.049) & 0.211 (0.047) & 0.229 (0.058) & 0.266 (0.075) \\
& 40 & 0.371 (0.118) &  0.316 (0.106) & 0.309 (0.106) & 0.339 (0.117) & 0.393 (0.112) \\
\hline
800 & 10 & 0.168 (0.029) & 0.162 (0.024)	& 0.162 (0.022)	& 0.164 (0.026)	 & 0.174 (0.036) \\
& 20 & 0.240 (0.069)	 & 0.210 (0.047) & 0.197 (0.042	) & 0.204 (0.046) & 0.233 (0.067) \\
& 40 & 0.351 (0.111)	& 0.280 (0.087)	 & 0.260 (0.070) & 0.292 (0.081) & 0.356 (0.092) \\
\hline
TCC \\
\hline
400 & 10 & 0.949 (0.033) & 0.959 (0.021) & 0.963 (0.016) & 0.961 (0.018)	 & 0.953 (0.029) \\
& 20 & 0.893 (0.053)	& 0.919 (0.034)	 & 0.932 (0.025) & 0.923 (0.039)  & 0.898	(0.059) \\
& 40 & 0.786 (0.108) & 0.834 (0.091) & 0.850 (0.083) & 0.829 (0.088) & 0.778 (0.104) \\
\hline
600 & 10 & 0.969 (0.015) & 0.972 (0.013) & 0.972 (0.013) & 0.970 (0.017) & 0.967 (0.016) \\
& 20 & 0.928 (0.048) & 0.953 (0.025) & 0.953 (0.025) & 0.944 (0.032) & 0.924 (0.046) \\
& 40 & 0.849 (0.101) & 0.889 (0.088) & 0.894 (0.090) & 0.871 (0.100) & 0.833 (0.099) \\
\hline
800 & 10 & 0.971 (0.010) & 0.973 (0.008) & 0.973 (0.007) & 0.972 (0.009)	 & 0.969 (0.014) \\
& 20 & 0.938 (0.040) & 0.954 (0.023) & 0.959 (0.019) & 0.956 (0.022) & 0.941 (0.039) \\
& 40 & 0.865 (0.089) & 0.914 (0.061) & 0.928 (0.046) & 0.908 (0.054) & 0.865 (0.070) \\
\hline
\end{tabularx}
\end{center}
\end{table}

\noindent \textbf{Example 2:} We now compare the performance of the transformed $\tau$-CQS with that of the linear $\tau$-CQS of Christou (2019).  The data is generated according to the following heteroscedastic models
\begin{eqnarray*}
\text{Model I}: \ \ Y=X_{1}+0.5X_{2}\varepsilon,\\
\text{Model II}: \ \ Y=X_{1}+0.5\exp(0.15X_{2})\varepsilon,\\
\text{Model III}: \ \ Y=X_{1}^3+0.5\exp(X_{2})\varepsilon,\\
\text{Model IV}: \ \ Y=\exp(X_{1})-1.05+0.5\exp(X_{2})\varepsilon,\\
\text{Model V}: \ \ Y=X_{1}^3+X_{2}+\frac{3\exp(2X_{3})}{1+\exp(2X_{3})}\varepsilon,\\
\text{Model VI}: \ \ Y=2\exp\left(\frac{X_{1}}{3} \right)+\frac{X_{2}^3}{3}+(X_{3}+X_{4})\varepsilon,
\end{eqnarray*}
where  $\mathbf{X}=(X_{1}, \dots, X_{10})^{\top}$ and the error $\varepsilon$ are generated according to a standard normal distribution.  
  
Table \ref{tab:example2a} reports the mean and standard deviation of DM and TCC for the two methods.  We observe that the transformed $\tau$-CQS outperforms the $\tau$-CQS for all models.  However, the performance of both  methods is comparable for Models I-IV and $\tau=0.5$, with $\tau$-CQS perform slightly better than the transformed $\tau$-CQS.  This is because for Models I-IV and $\tau=0.5$, the dimension of the transformed $\tau$-CQS and of the $\tau$-CQS is the same.    
  
\setlength{\arrayrulewidth}{.15em}
\begin{table}[h!]
\caption{\label{tab:example2a}\it{Mean (and standard deviation) of the estimation accuracy for $\widehat{\mathbf{T}}_{\tau}$ and $\widehat{\mathbf{B}}_{\tau}$, $\tau=0.1, 0.25, 0.5, 0.75, 0.9$, for Example 2.  TCQS denotes the proposed methodology and CQS denotes the linear $\tau$-CQS.  For each $\tau$, the value that is better in terms of estimation accuracy is bolded.}}
\begin{center}
\begin{tabularx}{\textwidth}{cc|@{\extracolsep\fill}ccccc}
\hline
DM \\
\hline
Model & Method & 0.1 & 0.25 & 0.5 & 0.75 & 0.9 \\
\hline
I & TCQS & \textbf{0.092} (0.029) & \textbf{0.071} (0.019) & 0.068 (0.018) & \textbf{0.071} (0.020) & \textbf{0.092} (0.027) \\
& CQS & 0.941 (0.071) & 0.937 (0.073) & \textbf{0.063} (0.017) & 0.934 (0.094) & 0.928 (0.098) \\
II & TCQS & \textbf{0.463} (0.021) & \textbf{0.279} (0.022) & 0.073 (0.019) & \textbf{0.282} (0.022) & \textbf{0.466} (0.021) \\
& CQS & 0.931 (0.086) & 0.938	 (0.078) & \textbf{0.068} (0.018) & 0.936 (0.076) & 0.937 (0.085) \\
III & TCQS & \textbf{0.357} (0.089) & \textbf{0.334} (0.079) & 0.342 (0.063) & \textbf{0.354} (0.076) & \textbf{0.372} (0.078) \\
& CQS & 0.932 (0.089) & 0.940 (0.083) & \textbf{0.301} (0.058) & 0.943 (0.082) & 0.941 (0.084) \\
IV & TCQS & \textbf{0.286} (0.079) & \textbf{0.309} (0.064) & 0.369 (0.062) & \textbf{0.405} (0.078) & \textbf{0.384} (0.086) \\
& CQS & 0.937 (0.075) & 0.939	 (0.064) & \textbf{0.322} (0.055) & 0.942 (0.070) & 0.941 (0.081) \\
V & TCQS & \textbf{0.480} (0.063) & \textbf{0.447} (0.076) & \textbf{0.410} (0.075) & \textbf{0.438} (0.074) & \textbf{0.488} (0.077) \\
& CQS & 0.968 (0.047) & 0.964	 (0.046) & 0.905 (0.109) & 0.965 (0.043) & 0.969 (0.047) \\
VI & TCQS &  \textbf{0.210} (0.061) & \textbf{0.187} (0.059) & \textbf{0.211} (0.061) & \textbf{0.192} (0.053) & \textbf{0.206} (0.059) \\
 & CQS & 0.961 (0.051) & 0.958 (0.058) & 0.942 (0.070) & 0.966 (0.047) & 0.964 (0.043) \\
\hline
TCC \\
\hline
Model & Method & 0.1 & 0.25 & 0.5 & 0.75 & 0.9 \\
\hline
I & TCQS & \textbf{0.991} (0.006) & \textbf{0.995} (0.003) & 0.995 (0.003) & \textbf{0.995} (0.003) & \textbf{0.991} (0.005) \\
& CQS & 0.714 (0.019) & 0.715 (0.018) & \textbf{0.996} (0.002) & 0.718 (0.029) & 0.720 (0.030) \\
II & TCQS & \textbf{0.785} (0.020) & \textbf{0.922} (0.012) & 0.994 (0.003) & \textbf{0.920} (0.012) & \textbf{0.782} (0.020) \\
& CQS & 0.716 (0.025) & 0.715 (0.023) & \textbf{0.995} (0.002) & 0.715 (0.023) & 0.715 (0.025) \\
III & TCQS & \textbf{0.865} (0.066) & \textbf{0.882} (0.054) & 0.879 (0.043) & \textbf{0.869} (0.053) & \textbf{0.856} (0.058)\\
& CQS & 0.710 (0.027) & 0.710  (0.026) & \textbf{0.906} (0.034) & 0.709 (0.024) & 0.707 (0.027)\\
IV & TCQS & \textbf{0.912} (0.048) & \textbf{0.901} (0.040) & 0.860 (0.046) & \textbf{0.830} (0.064) & \textbf{0.845} (0.0690 \\
& CQS & 0.700 (0.024) & 0.703 (0.017) & \textbf{0.893} (0.035) & 0.708 (0.019) & 0.709 (0.024) \\
V & TCQS & \textbf{0.766} (0.061) & \textbf{0.795} (0.070) & \textbf{0.827} (0.063) & \textbf{0.803} (0.066) & \textbf{0.756} (0.077) \\
& CQS & 0.637 (0.049) & 0.639	 (0.048) & 0.720 (0.035) & 0.633 (0.048) & 0.629 (0.050) \\
VI & TCQS & \textbf{0.952} (0.030) & \textbf{0.961} (0.023) & \textbf{0.952} (0.027) & \textbf{0.960} (0.022) & \textbf{0.954} (0.026) \\
 & CQS &  0.630 (0.044) & 0.630 (0.048) & 0.702 (0.020) & 0.632 (0.044) & 0.633 (0.047) \\
\hline
\end{tabularx}
\end{center}
\end{table}

\noindent \textbf{Example 3:} We further compare the performance of the transformed $\tau$-CQS with that of the linear $\tau$-CQS using different distributions for $\mathbf{X}$.  Specifically, we consider an $\mathbf{X}$ with dependent components, i.e., $\mathbf{X}=(X_{1}, \dots, X_{10})^{\top} \sim N(\mathbf{0}, (\sigma_{ij})_{1 \leq i,j \leq 10})$ with $\sigma_{ij}=0.5^{|i-j|}$, and also an $\mathbf{X}$ that follows a $t$-distribution with 3, 5, or 10 degrees of freedom.  To save space, and since the results follow similar pattern, we only report the results for Model I.  Table \ref{tab:example2b} demonstrates the mean and standard deviation of DM and TCC for the two methods and the different distributions.  We observe that the estimation accuracy for $\mathbf{X}$ with dependent components is smaller than that for $\mathbf{X}$ with independent components.  However, the degree over the transformed $\tau$-CQS improves upon $\tau$-CQS is the same.  Moreover, the estimation accuracy for $\mathbf{X}$ with a $t$ distribution improves with increasing degrees of freedom.  

\setlength{\arrayrulewidth}{.15em}
\begin{table}[h!]
\caption{\label{tab:example2b}\it{Mean (and standard deviation) of the estimation accuracy for $\widehat{\mathbf{T}}_{\tau}$ and $\widehat{\mathbf{B}}_{\tau}$, $\tau=0.1, 0.25, 0.5, 0.75, 0.9$, for Example 3.  TCQS denotes the proposed methodology and CQS denotes the linear $\tau$-CQS.  For each $\tau$, the value that is better in terms of estimation accuracy is bolded.}}
\begin{center}
\begin{tabularx}{\textwidth}{cc|@{\extracolsep\fill}ccccc}
\hline
DM \\
\hline
 Method & $\mathbf{X}$ & 0.1 & 0.25 & 0.5 & 0.75 & 0.9 \\
\hline
TCQS & Normal & \textbf{0.257} (0.069) & \textbf{0.219} (0.052) & 0.192 (0.037) & \textbf{0.216} (0.056) & \textbf{0.252} (0.074) \\
 & $t_{3}$ & \textbf{0.431} (0.144) & \textbf{0.418} (0.155) & 0.381 (0.135) & \textbf{0.254} (0.111) & \textbf{0.207} (0.102)  \\
 & $t_{5}$ & \textbf{0.270} (0.085) & \textbf{0.233} (0.089) & 0.186 (0.070) & \textbf{0.142} (0.056) & \textbf{0.138} (0.062) \\
 & $t_{10}$ & \textbf{0.191} (0.058) & \textbf{0.140} (0.043) & 0.103 (0.028) & \textbf{0.091} (0.026) & \textbf{0.096} (0.032) \\
 CQS & Normal & 0.948 (0.072) & 0.939 (0.080) & \textbf{0.155} (0.043) & 0.927 (0.078) & 0.940 (0.0770 \\
 & $t_{3}$ & 0.933 (0.080) & 0.937 (0.071) & \textbf{0.226} (0.109) & 0.930 (0.088) & 0.952 (0.064) \\
 & $t_{5}$ & 0.921 (0.101) & 0.918 (0.118) & \textbf{0.138} (0.048) & 0.944 (0.075) & 0.959 (0.059) \\
 & $t_{10}$ & 0.932 (0.077) & 0.919 (0.092) & \textbf{0.102} (0.033) & 0.946 (0.065) & 0.959 (0.053) \\
\hline
TCC \\
\hline
 Method & $\mathbf{X}$ & 0.1 & 0.25 & 0.5 & 0.75 & 0.9 \\
\hline
 TCQS & Normal & \textbf{0.929} (0.038) & \textbf{0.949} (0.024) & 0.962 (0.014) & \textbf{0.950} (0.026) & \textbf{0.931} (0.041) \\
 & $t_{3}$ & \textbf{0.793} (0.138) & \textbf{0.801} (0.143) & 0.837 (0.116) & \textbf{0.923} (0.078) & \textbf{0.947} (0.063) \\
 & $t_{5}$ & \textbf{0.920} (0.051) & \textbf{0.938} (0.049) & 0.961 (0.032) & \textbf{0.977} (0.021) & \textbf{0.977} (0.022)\\
 & $t_{10}$ & \textbf{0.960} (0.025) & \textbf{0.979} (0.013) & \textbf{0.989} (0.007) & \textbf{0.991} (0.005) & \textbf{0.990} (0.007) \\
  CQS & Normal & 0.701 (0.020) & 0.705 (0.024) & \textbf{0.974} (0.015) & 0.706 (0.023) & 0.704 (0.022) \\
& $t_{3}$ & 0.695 (0.028) & 0.696 (0.025) & \textbf{0.937} (0.077) & 0.705 (0.029) & 0.698 (0.021) \\
& $t_{5}$ & 0.713 (0.033) & 0.716 (0.041) & \textbf{0.979} (0.014) & 0.710 (0.021) & 0.706 (0.014) \\
 & $t_{10}$ & 0.713 (0.023) & 0.718 (0.029) & \textbf{0.989} (0.009) & 0.711 (0.015) & 0.708 (0.013)\\
\hline
\end{tabularx}
\end{center}
\end{table}

\begin{figure}[h!] 
\begin{center}
\includegraphics[width=\textwidth]{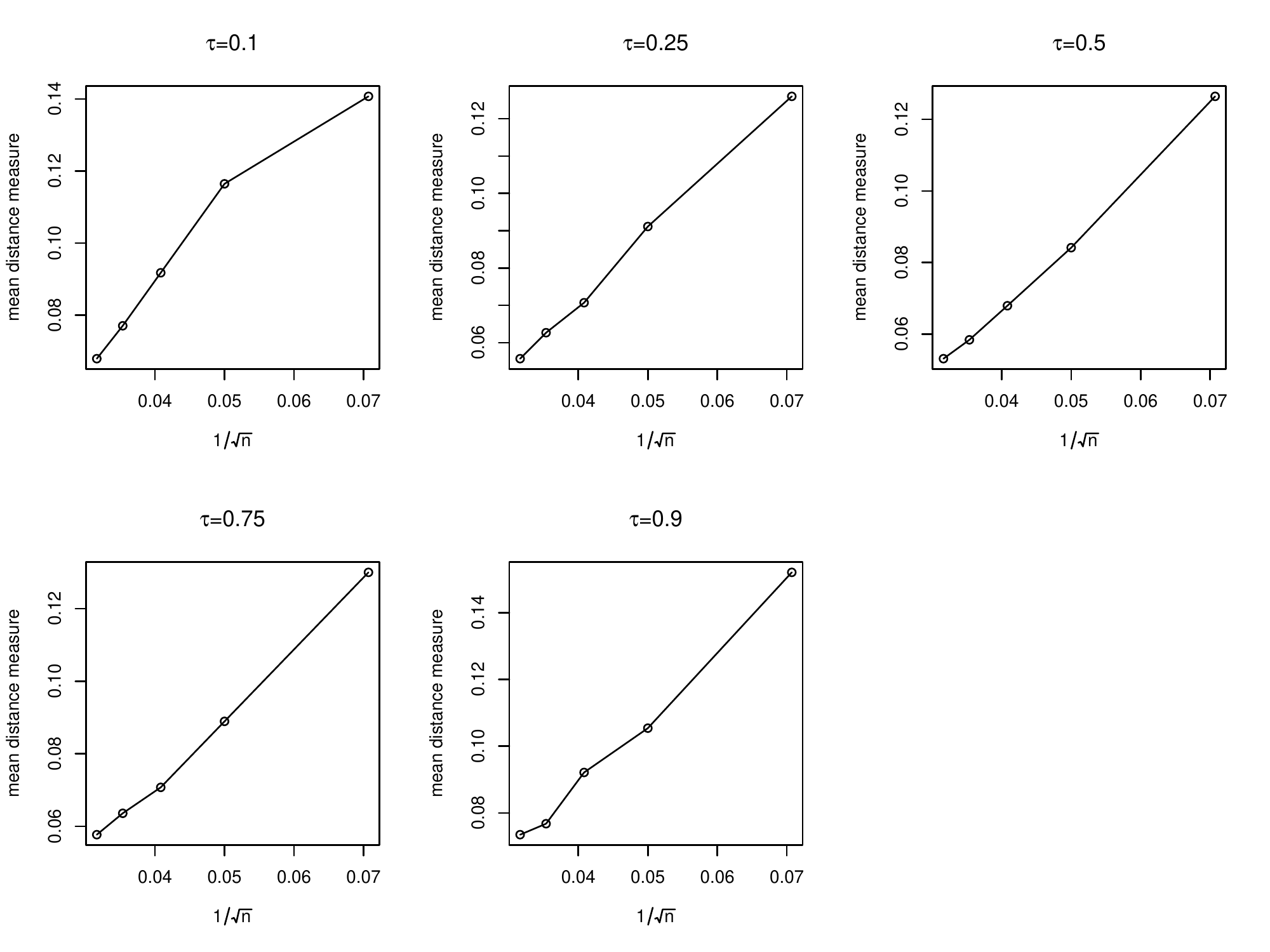} \\ 
\caption{\label{fig:example4}{\it{The $\sqrt{n}$-consistency of the proposed estimator $\widehat{\mathbf{T}}_{\tau}$ for Example 4.}}}
\end{center}
\end{figure}

\noindent \textbf{Example 4:}  Finally, we demonstrate the $\sqrt{n}$-consistency of the proposed methodology, stated in Theorem \ref{thm1}.  We reconsider Model I, where $\mathbf{X}=(X_{1}, \dots, X_{10})^{\top}$ and the error $\varepsilon$ are generated according to a standard normal distribution.  The sample size is taken to be $n=400, 600, \dots, 1200$.  Due to space limitation and since the results show similar pattern, we only present the mean DM.  Figure \ref{fig:example4} indicates an approximate linear relationship between the mean DM and $1/\sqrt{n}$, demonstrating the $\sqrt{n}$-consistency of the proposed estimator.       

\subsection{Real Data Analysis}

\subsubsection{Vowel Recognition}  

This data set consists of measurements on 11 variables.  The dependent variable of interest is a categorial variable with 11 levels, representing different vowel sounds, and the other 10 variables describe the features of a sound.  The data can be found in the UCI Machine Learning Repository (http://archive.ics.uci.edu/ml/datasets.html).  Here we focus on only three vowels: the sounds in heed, head and hud. 

Li et al. (2011) considered this data set and concluded that the kernel principal support vector machine achieves much better separation of the three vowels than other linear dimension reduction techniques, such as SIR, SAVE, and DR.  However, those methods are focusing on the linear and nonlinear CS.  We investigate the transformed $\tau$-CQS for different values of $\tau$.  

The data set is separated into training and testing sets, which have sample sizes of 144 and 126, respectively.  We use the training set to find the estimated vectors for the transformed $\tau$-CQS and we evaluate them at the test set.  Figure \ref{fig:Real1} (a) presents the first two column vectors of $\widehat{\mathbf{T}}_{\tau}^{\top}\widehat{\mathbf{g}}(\mathbf{X}_{test})$, for $\tau=0.1, 0.25, 0.5, 0.75$ and 0.9.  The plots show strong separation of the three vowels and demonstrate the necessity of different quantile levels. 

For further comparisons we calculate the correlation between the response variable and the first two column vectors of $\widehat{\mathbf{T}}_{\tau}^{\top}\widehat{\mathbf{g}}(\mathbf{X}_{test})$ and $\widehat{\mathbf{B}}_{\tau}^{\top}\mathbf{X}_{test}$.  From Table \ref{tab:Real1} (a) we observe that the estimated transformed sufficient predictors $\widehat{\mathbf{T}}_{\tau}^{\top}\widehat{\mathbf{g}}(\mathbf{X}_{test})$ explain more variability of the response than that explained by the linear sufficient predictors $\widehat{\mathbf{B}}_{\tau}^{\top}\mathbf{X}_{test}$. 

\subsubsection{Breast Cancer Diagnostic}

This data set consists of measurements on 10 variables.  The dependent variable of interest is a categorical variable indicating whether the diagnosis is benign or malignant, and the other 9 variables describe characteristics of the cell.  The data can be found in the UCI Machine Learning Repository (http://archive.ics.uci.edu/ml/datasets.html). 

\begin{figure}[h!] 
\begin{center}
%$\tau=0.1$ \hspace{2cm} $\tau=0.25$ \hspace{2cm} $\tau=0.5$ \hspace{2cm} $\tau=0.75$ \hspace{2cm} $\tau=0.9$ \\
\includegraphics[width=\textwidth]{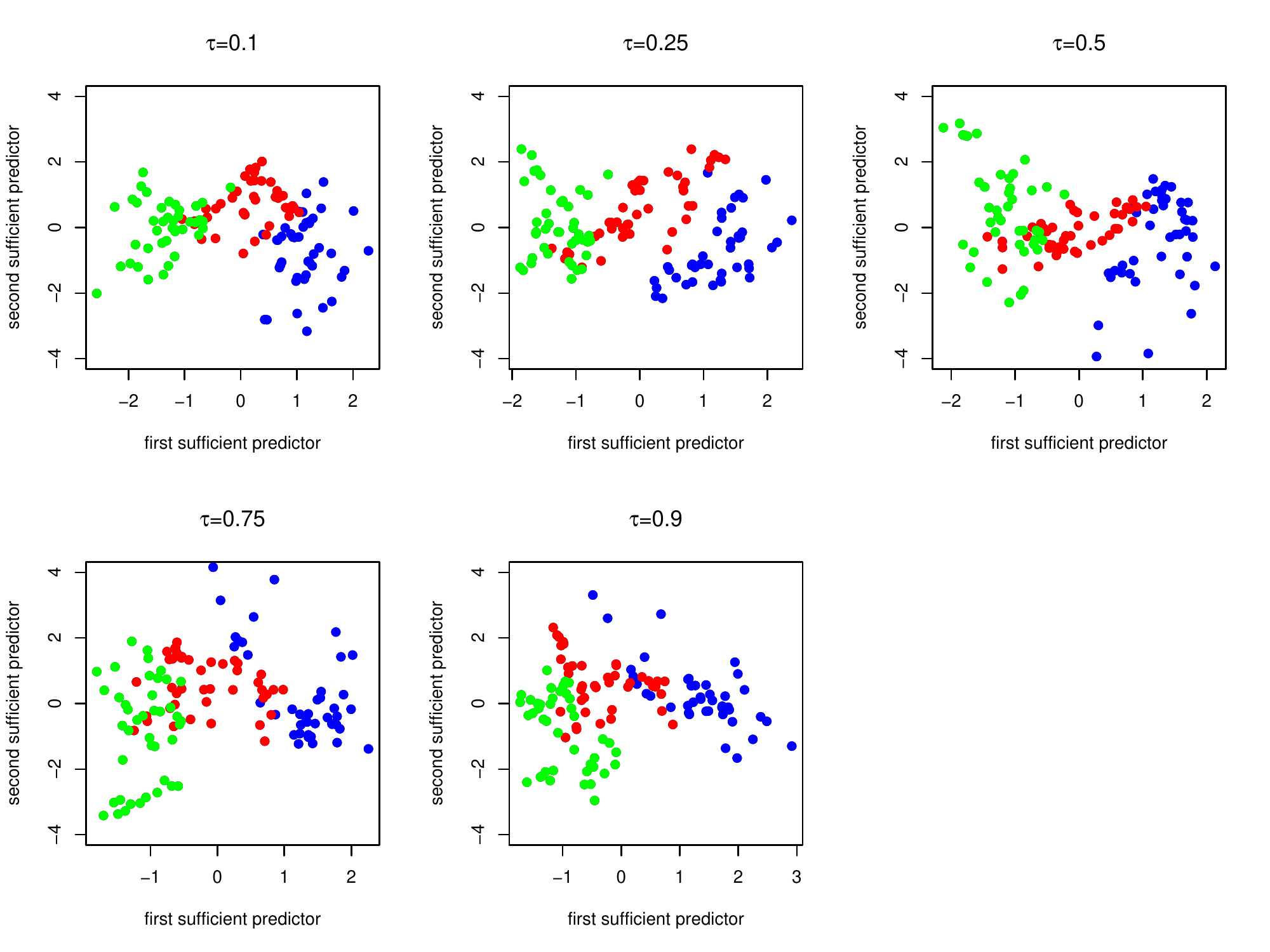} \\ (a) \\ \includegraphics[width=\textwidth]{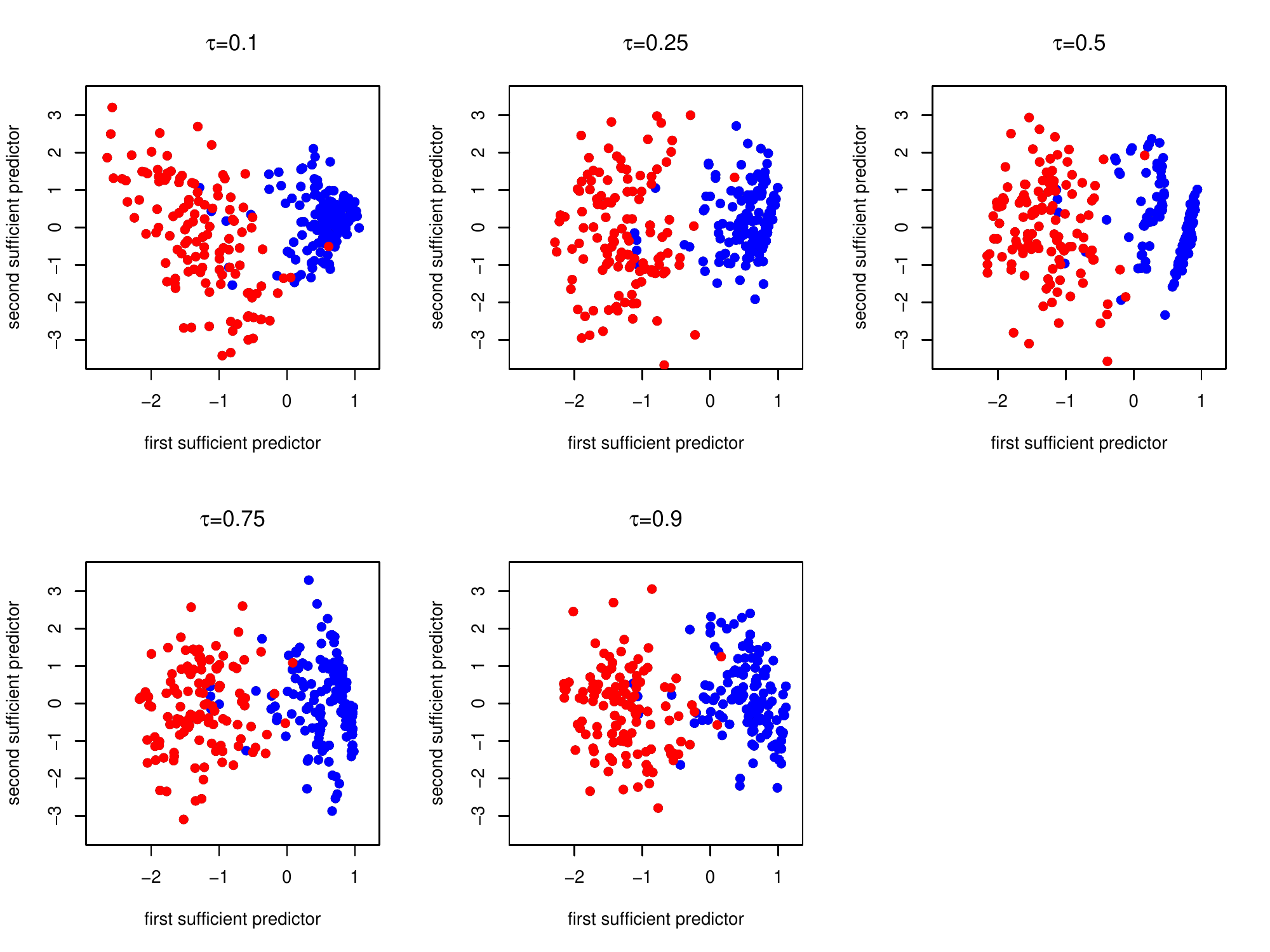} \\ (b)
\caption{\label{fig:Real1}{\it{First two sufficient predictors for the transformed $\tau$-CQS, for $\tau=0.1, 0.25, 0.5, 0.75$ and 0.9.  (a) Blue, red, and green colors indicate the vowel sounds in heed, head, and hud, respectively. (b) Blue and red colors indicate benign and malignant, respectively.}}}
\end{center}
\end{figure}

The data set consists of $n=682$ observations.  We randomly divide the data set into two halves, representing a training set and a test set.  Figure \ref{fig:Real1} (b) presents the first two column vectors of $\widehat{\mathbf{T}}_{\tau}^{\top}\widehat{\mathbf{g}}(\mathbf{X}_{test})$, for $\tau=0.1, 0.25, 0.5, 0.75$ and 0.9.  The plots show strong separation of the two classes of diagnosis.  Moreover, Table \ref{tab:Real1} (b) shows that the estimated transformed sufficient predictors $\widehat{\mathbf{T}}_{\tau}^{\top}\widehat{\mathbf{g}}(\mathbf{X}_{test})$ explain more variability of the response, especially the variability explained by the second estimated transformed sufficient predictor.  
    
\setlength{\arrayrulewidth}{.15em}
\begin{table}[h!]
\caption{\label{tab:Real1}\it{Correlation between the response and the first two sufficient predictors obtained by the transformed $\tau$-CQS and the linear $\tau$-CQS, for $\tau=0.1, 0.25, 0.5, 0.75, 0.9$.  The first line represents the correlation with the first sufficient predictor and the second line represents the correlation with the second sufficient predictor.  For each $\tau$, the value with the largest, in absolute value, correlation is bolded.}}
\begin{center}
\begin{tabularx}{\textwidth}{cc|@{\extracolsep\fill}rrrrr}
\hline
Application & Direction & 0.1 & 0.25 & 0.5 & 0.75 & 0.9 \\
\hline
(a) & \\
\hline
Transformed & dir1 & \textbf{-0.939} & -0.927 & \textbf{-0.928} & \textbf{-0.927} & \textbf{-0.905} \\
& dir2 & \textbf{0.599} & \textbf{0.664} & \textbf{0.761} & \textbf{-0.663} & \textbf{-0.635} \\
Linear & dir1 & -0.933 & \textbf{-0.932} & -0.925 & -0.896 & -0.880 \\
& dir2 & 0.388 & 0.433 & 0.544 & 0.450 & -0.376 \\
\hline
(b) & \\
\hline
Transformed & dir1 & -0.873 & \textbf{-0.902} & -0.900 & -0.901 & \textbf{-0.902} \\
& dir2 & \textbf{-0.403} & \textbf{-0.098} & \textbf{0.424} & \textbf{0.113} & 0.032 \\
Linear & dir1 & \textbf{-0.894} & -0.900 & \textbf{0.903} & \textbf{-0.905} & -0.895 \\
& dir2 & 0.132 & 0.068 & -0.014 & -0.067 & \textbf{-0.137} \\
\hline
\end{tabularx}
\end{center}
\end{table}

\section{Discussion} \label{Disc}

In this work we have considered the transformed dimension reduction for conditional quantiles, which serves as an intermediate step between linear and nonlinear dimension reduction for conditional quantiles.  The idea is a straightforward extension of Wang et al. (2014)'s methodology and it considers transforming the predictors monotonically and then apply linear dimension reduction in the space defined by the transformed variables.  Simulation examples and real data applications demonstrate the performance of the proposed methodology and show the degree by which it outperforms the linear $\tau$-CQS.  

\appendix

\section{Notation and Assumptions}\label{AppendixA}

\noindent \underline{\textbf{Notation}}
\begin{enumerate}
\item[N1] Recall that,
\begin{enumerate}
\item[(a)] $\mathbf{g}(\mathbf{X})=(g_{1}(X_{1}), \dots, g_{p}(X_{p}))^{\top}=(\Phi^{-1}\{F_{1}(X_{1})\}, \dots, \Phi^{-1}\{F_{p}(X_{p})\})^{\top}$, where $\Phi(\cdot)$ and $F_{j}(\cdot)$ denote the standard normal distribution function and the marginal distribution function of $X_{j}$, respectively.
\item[(b)] $\widehat{\mathbf{g}}(\mathbf{X})=(\widehat{g}_{1}(X_{1}), \dots, \widehat{g}_{p}(X_{p}))^{\top}=(\Phi^{-1}\{\widehat{F}_{1}(X_{1})\}, \dots, \Phi^{-1}\{\widehat{F}_{p}(X_{p})\})$, where $\widehat{F}_{j}(\cdot)$ denotes the empirical marginal distribution function of $X_{j}$. 
\end{enumerate}

\item[N2]  We say that a function $m(\cdot): \mathbb{R}^{p} \rightarrow \mathbb{R}$ has the order of smoothness $s$ on the support $\mathcal{X}$, denoted by $m(\cdot) \in H_{s}(\mathcal{X})$, if
\begin{enumerate}
\item[(a)] it is differentiable up to order $[s]$, where $[s]$ denotes the lowest integer part of $s$, and 
\item[(b)] there exists a constant $L>0$, such that for all $\mathbf{u}=(u_{1}, \dots, u_{p})^{\top}$ with $|\mathbf{u}|=u_{1}+ \cdots + u_{p}=[s]$, all $\tau$ in an interval $[\underline{\tau}, \bar{\tau}]$, where $0 < \underline{\tau} \leq \bar{\tau} <1$, and all $\mathbf{x}, \mathbf{x}'$ in $\mathcal{X}$,
\begin{eqnarray*}
|D^{\mathbf{u}}m(\mathbf{x}) - D^{\mathbf{u}}m(\mathbf{x}')| \leq L \left\| \mathbf{x}-\mathbf{x}' \right\|^{s-[s]},
\end{eqnarray*}
where $D^{\mathbf{u}}m(\mathbf{x})$ denotes the partial derivative $\partial^{|\mathbf{u}|}m(\mathbf{x})/ \partial x_{1}^{u_{1}} \dots x_{p}^{u_{p}}$ and $\left\| \cdot \right\|$ denotes the Euclidean norm. 
\end{enumerate}
\end{enumerate}

\noindent \underline{\textbf{Assumptions}}
\begin{enumerate}
\item[A1] The following moment conditions are satisfied
\begin{eqnarray*}
E \left\| \mathbf{g}(\mathbf{X}) \mathbf{g}^{\top}(\mathbf{X}) \right\| < \infty,  \ \ E|Q_{\tau}\{Y|\mathbf{\Gamma}^{\top} \mathbf{g}(\mathbf{X})\}|^{2} < \infty,  \ \ E\left[Q_{\tau}\{Y|\mathbf{\Gamma}^{\top}\mathbf{g}(\mathbf{X})\}^{2} \left\| \mathbf{g}(\mathbf{X})\mathbf{g}^{\top}(\mathbf{X}) \right\| \right] < \infty,
\end{eqnarray*}
for a given $\tau \in (0,1)$.  

\item[A2] The distribution of $\mathbf{\Gamma}^{\top} \mathbf{g}(\mathbf{X})$ has a probability density function $f_{\mathbf{\Gamma}}(\cdot)$ with respect to the Lebesgue measure, which is strictly positive and continuously differentiable over the support $\mathcal{X}_{\mathbf{g}}$ of $\mathbf{g}(\mathbf{X})$.  

\item[A3] The cumulative distribution function $F_{Y|\mathbf{\Gamma}}(\cdot | \cdot)$ of $Y$ given $\mathbf{\Gamma}^{\top} \mathbf{g}(\mathbf{X})$ has a continuous probability density function $f_{Y|\mathbf{\Gamma}} \{y|\mathbf{\Gamma}^{\top}\mathbf{g}(\mathbf{x})\}$ with respect to the Lebesgue measure, which is strictly positive for $y \in \mathbb{R}$ and $\mathbf{\Gamma}^{\top}\mathbf{g}(\mathbf{x})$, for $\mathbf{g}(\mathbf{x}) \in \mathcal{X}_{\mathbf{g}}$.  The partial derivative $\partial F_{Y|\mathbf{\Gamma}}\{y|\mathbf{\Gamma}^{\top}\mathbf{g}(\mathbf{x})\}/ \partial \mathbf{\Gamma}^{\top}\mathbf{g}(\mathbf{x})$ is continuous.  There is a $L_{0}>0$, such that
\begin{eqnarray*}
|f_{Y|\mathbf{\Gamma}}\{y|\mathbf{\Gamma}^{\top}\mathbf{g}(\mathbf{x})\}-f_{Y|\mathbf{\Gamma}}\{y'|\mathbf{\Gamma}^{\top}\mathbf{g}(\mathbf{x}')\}| \leq L_{0} \left\| (\mathbf{\Gamma}^{\top}\mathbf{g}(\mathbf{x}),y)-(\mathbf{\Gamma}^{\top}\mathbf{g}(\mathbf{x}'),y') \right\| \ 
\end{eqnarray*}
for all $(\mathbf{g}(\mathbf{x}),y), (\mathbf{g}(\mathbf{x}'),y') \ \text{of} \ \mathcal{X}_{\mathbf{g}} \times \mathbb{R}$.

\item[A4] The nonnegative kernel function $K(\cdot)$, used in (\ref{llqr}), is Lipschitz over $\mathbb{R}^{d^*}$, where $d^*$ is the dimension of $K(\cdot)$, and satisfies $\int K(\mathbf{z})d \mathbf{z}=1$.  For some $\underline{K}>0$, $K(\mathbf{z}) \geq \underline{K} I\{\mathbf{z} \in B(0,1)\}$, where $B(0,1)$ is the closed unit ball.  The associated bandwidth $h$, used in the estimation procedure, is in $[\underline{h},\overline{h}]$ with $0< \underline{h} \leq \overline{h} < \infty$, $\lim_{n \rightarrow \infty} \overline{h}=0$ and $\lim_{n \rightarrow \infty} (\ln{n})/(n \underline{h}^{d^*})=0$.

\item[A5] $Q_{\tau}\{Y|\mathbf{\Gamma}^{\top}\mathbf{g}(\mathbf{x})\}$ is in $H_{s_{\tau}}(\mathcal{T}_{\mathbf{\Gamma}})$ for some $s_{\tau}$ with $[s_{\tau}] \leq 1$, where $\mathcal{T}_{\mathbf{\Gamma}}=\{\mathbf{z} \in \mathbb{R}^{d_{2}}: \mathbf{z}=\mathbf{\Gamma}^{\top}\mathbf{g}(\mathbf{x}), \mathbf{g}(\mathbf{x}) \in \mathcal{X}_{\mathbf{g}}\}$, and $\mathcal{X}_{\mathbf{g}}$ is the support of $\mathbf{g}(\mathbf{X})$. 
\end{enumerate}

Assumptions A2-A4 come from the work of Guerre and Sabbah (2012) and are necessary for the uniform consistency of $\widehat{Q}_{\tau}\{Y|\widehat{\mathbf{\Gamma}}^{\top}\widehat{\mathbf{g}}(\mathbf{x})\}$ defined in connection with (\ref{llqr}).  

\section{Proof of Main Results}\label{AppendixB}
\renewcommand{\theequation}{B.\arabic{equation}}
\renewcommand{\thethm}{B.\arabic{thm}}

\subsection{Some Lemmas}\label{Lemmas}

\begin{lemma}\label{LemmaB1}
Under Assumptions A2-A5 given in Appendix \ref{AppendixA}, and the assumption that $\widehat{\mathbf{\Gamma}}$ is $\sqrt{n}$-consistent estimate of the directions of the transformed CS, then
\begin{eqnarray*}
\sup_{\mathbf{x} \in \mathcal{X}} |\widehat{Q}_{\tau}\{Y|\widehat{\mathbf{\Gamma}}^{\top}\widehat{\mathbf{g}}(\mathbf{x})\}-Q_{\tau}\{Y|\mathbf{\Gamma}^{\top}\mathbf{g}(\mathbf{x})\}|=O_{p}(1),
\end{eqnarray*}
where $\mathcal{X}$ denotes the support of $\mathbf{X}$, and $\widehat{Q}_{\tau}\{Y|\widehat{\mathbf{\Gamma}}^{\top}\mathbf{g}(\mathbf{x})\}$ denotes the local linear conditional quantile estimate of $Q_{\tau}\{Y|\mathbf{\Gamma}^{\top}\mathbf{g}(\mathbf{x})\}$, defined in connection with (\ref{llqr}).  
\end{lemma}

\noindent \textbf{Proof:}
Observe that
\begin{eqnarray*}
&& \sup_{\mathbf{x} \in \mathcal{X}} |\widehat{Q}_{\tau}\{Y|\widehat{\mathbf{\Gamma}}^{\top}\widehat{\mathbf{g}}(\mathbf{x})\}-Q_{\tau}\{Y|\mathbf{\Gamma}^{\top}\mathbf{g}(\mathbf{x})\}|\\
 &\leq& \sup_{\mathbf{x} \in \mathcal{X}} |\widehat{Q}_{\tau}\{Y|\widehat{\mathbf{\Gamma}}^{\top}\widehat{\mathbf{g}}(\mathbf{x})\}-\widehat{Q}_{\tau}\{Y|\mathbf{\Gamma}^{\top}\mathbf{g}(\mathbf{x})\}|\\
&&+\sup_{\mathbf{x} \in \mathcal{X}} |\widehat{Q}_{\tau}\{Y|\mathbf{\Gamma}^{\top}\mathbf{g}(\mathbf{x})\}-Q_{\tau}\{Y|\mathbf{\Gamma}^{\top}\mathbf{g}(\mathbf{x})\}|\\
&=&O_{p}(1).
\end{eqnarray*}
The first term follows from the Bahadur representation of $\widehat{Q}_{\tau}\{Y|\widehat{\mathbf{\Gamma}}^{\top}\widehat{\mathbf{g}}(\mathbf{x})\}-\widehat{Q}_{\tau}\{Y|\mathbf{\Gamma}^{\top}\mathbf{g}(\mathbf{x})\}$ (see Guerre and Sabbah 2012) and the $\sqrt{n}$-consistency of $\widehat{\mathbf{\Gamma}}$.  The second term follows from Corollary 1 (ii) of Guerre and Sabbah (2012).

\begin{lemma}\label{LemmaB2}
For a given $\tau \in (0,1)$, assume that $Y \independent Q_{\tau} \{Y|\mathbf{g}(\mathbf{X})\} | \mathbf{T}_{\tau}^{\top} \mathbf{g}(\mathbf{X})$, where $\mathbf{T}_{\tau}$ is a $p \times 1$ vector.  If $\mathbf{g}(\mathbf{X}) \sim N(\mathbf{0}, \mathbf{\Sigma}_{\mathbf{g}})$, and Assumptions A1-A5, given in Appendix A, hold, then $\widehat{\boldsymbol{\eta}}_{\tau}$ is $\sqrt{n}$-consistent estimate of the direction of $\mathcal{S}_{Q_{\tau}\{Y|\mathbf{g}(\mathbf{X})\}}$, where $\widehat{\boldsymbol{\eta}}_{\tau}$ is defined in (\ref{OLS_transformed}).  
\end{lemma}

\noindent \textbf{Proof:} Observe that minimizing $\sum_{i=1}^{n}[\widehat{Q}_{\tau}\{Y|\widehat{\mathbf{\Gamma}}^{\top}\widehat{\mathbf{g}}(\mathbf{X}_{i})\}-\gamma_{\tau}-\boldsymbol{\eta}_{\tau}^{\top}\widehat{\mathbf{g}}(\mathbf{X}_{i})]^2$ with respect to $(\gamma_{\tau},\boldsymbol{\eta}_{\tau})$, is equivalent with minimizing 
\begin{eqnarray}\label{Snhat}
\widehat{S}_{n}(\gamma_{\tau},\boldsymbol{\eta}_{\tau})&=&\frac{1}{2}\sum_{i=1}^{n}[\widehat{Q}_{\tau}\{Y|\widehat{\mathbf{\Gamma}}^{\top}\widehat{\mathbf{g}}(\mathbf{X}_{i})\}-\gamma_{\tau}-\boldsymbol{\eta}_{\tau}^{\top}\widehat{\mathbf{g}}(\mathbf{X}_{i})]^2 -\frac{1}{2}\sum_{i=1}^{n}[\widehat{Q}_{\tau}\{Y|\widehat{\mathbf{\Gamma}}^{\top}\widehat{\mathbf{g}}(\mathbf{X}_{i})\}]^2 \nonumber \\
&=&-(\gamma_{\tau},\boldsymbol{\eta}^{\top}_{\tau})^{\top}\sum_{i=1}^{n}\widehat{Q}_{\tau}\{Y|\widehat{\mathbf{\Gamma}}^{\top}\widehat{\mathbf{g}}(\mathbf{X}_{i})\}(1,\widehat{\mathbf{g}}(\mathbf{X}_{i})) \nonumber \\
&&+\frac{1}{2}(\gamma_{\tau},\boldsymbol{\eta}^{\top}_{\tau})^{\top}\sum_{i=1}^{n}(1,\widehat{\mathbf{g}}(\mathbf{X}_{i}))(1,\widehat{\mathbf{g}}^{\top}(\mathbf{X}_{i}))^{\top}(\gamma_{\tau},\boldsymbol{\eta}_{\tau})
\end{eqnarray}
with respect to $(\gamma_{\tau},\boldsymbol{\eta}_{\tau})$.

Let $\widehat{S}_{n}(\boldsymbol{c}_{\tau}/\sqrt{n}+(\gamma^*_{\tau},\boldsymbol{\eta}^*_{\tau}))$ be as defined in (\ref{Snhat}), where $\boldsymbol{c}_{\tau}=\sqrt{n} \{(\gamma_{\tau},\boldsymbol{\eta}_{\tau})-(\gamma^*_{\tau},\boldsymbol{\eta}^*_{\tau})\}$ and $(\gamma^*_{\tau}, \boldsymbol{\eta}_{\tau}^*)$ satisfies $(\gamma_{\tau}^*, \boldsymbol{\eta}^*_{\tau})= \arg \min_{(\gamma_{\tau},\boldsymbol{\eta}_{\tau})} E[Q_{\tau}\{Y|\mathbf{\Gamma}^{\top}\mathbf{g}(\mathbf{X})\}-\gamma_{\tau}-\boldsymbol{\eta}_{\tau}^{\top}\mathbf{g}(\mathbf{X})]^2$.
Assume that the following quadratic approximation holds,  uniformly in $\mathbf{c}_{\tau}$ in a compact set, 
\begin{eqnarray}\label{quadr_approx}
\widehat{S}_{n}(\boldsymbol{c}_{\tau}/\sqrt{n}+(\gamma^*_{\tau},\boldsymbol{\eta}^*_{\tau}))= \frac{1}{2}\boldsymbol{c}_{\tau}^{\top}\mathbb{V}\boldsymbol{c}_{\tau}+\mathbf{W}_{\tau,n}^{\top}\boldsymbol{c}_{\tau}+C_{\tau,n}+o_{p}(1),
\end{eqnarray}
where $\mathbb{V}=E\{(1,\mathbf{g}(\mathbf{X}))(1,\mathbf{g}^{\top}(\mathbf{X}))^{\top}\}$, 
\begin{eqnarray} \label{Wn}
\mathbf{W}_{\tau,n}=-\frac{1}{\sqrt{n}}\sum_{i=1}^{n} \widehat{Q}_{\tau}\{Y|\widehat{\mathbf{\Gamma}}^{\top}\widehat{\mathbf{g}}(\mathbf{X}_{i})\}(1,\widehat{\mathbf{g}}(\mathbf{X}_{i})),
\end{eqnarray}
and
\begin{eqnarray} \label{Cn}
C_{\tau,n}=-\sum_{i=1}^{n}\widehat{Q}_{\tau}\{Y|\widehat{\mathbf{\Gamma}}^{\top}\widehat{\mathbf{g}}(\mathbf{X}_{i})\}(1,\widehat{\mathbf{g}}^{\top}(\mathbf{X}_{i}))^{\top}(\gamma^*_{\tau},\boldsymbol{\eta}^*_{\tau}) +\frac{1}{2}(\gamma^*_{\tau},\boldsymbol{\eta}^{*\top}_{\tau})^{\top}\sum_{i=1}^{n}(1,\widehat{\mathbf{g}}(\mathbf{X}_{i}))(1,\widehat{\mathbf{g}}^{\top}(\mathbf{X}_{i}))^{\top}(\gamma^*_{\tau},\boldsymbol{\eta}^*_{\tau}).
\end{eqnarray}

Then, to prove the $\sqrt{n}$-consistency of $\widehat{\boldsymbol{\eta}}_{\tau}$, enough to show that for any given $\delta_{\tau}>0$, there exists a constant $\Lambda_{\tau}$ such that
\begin{eqnarray}\label{positiveprobability}
\Pr  \left \{ \inf_{\left\|\boldsymbol{c}_{\tau} \right\| \geq \Lambda_{\tau}}  \widehat{S}_{n}(\boldsymbol{c}_{\tau}/\sqrt{n}+(\gamma^*_{\tau},\boldsymbol{\eta}^*_{\tau}))>\widehat{S}_{n}(\gamma^*_{\tau},\boldsymbol{\eta}^*_{\tau}) \right \} \geq 1-\delta_{\tau}.
\end{eqnarray}
This implies that with probability at least $1-\delta_{\tau}$ there exists a local minimum in the ball $\{\boldsymbol{c}_{\tau}/\sqrt{n}+(\gamma_{\tau}^*,\boldsymbol{\eta}^*_{\tau}): \left\| \boldsymbol{c}_{\tau} \right\| \leq \Lambda_{\tau}\}$.  This in turn implies that there exists a local minimizer such that $\left\|(\widehat{\gamma}_{\tau},\widehat{\boldsymbol{\eta}}_{\tau})-(\gamma_{\tau}^*,\boldsymbol{\eta}^*_{\tau}) \right\|=O_{p}\left(n^{-1/2} \right)$. From (\ref{quadr_approx})
\begin{eqnarray}\label{Difference}
\widehat{S}_{n}(\boldsymbol{c}_{\tau}/\sqrt{n}+(\gamma_{\tau}^*,\boldsymbol{\eta}^*_{\tau}))-\widehat{S}_{n}(\gamma_{\tau}^*,\boldsymbol{\eta}^*_{\tau})=\frac{1}{2}\boldsymbol{c}_{\tau}^\top \mathbb{V}\boldsymbol{c}_{\tau}+\mathbf{W}^\top_{\tau,n}\boldsymbol{c}_{\tau}+o_{p}(1),
\end{eqnarray}
for any $\boldsymbol{c}_{\tau}$ in a compact subset of $\mathbb{R}^{p+1}$.  Therefore, the difference (\ref{Difference}) is dominated by the quadratic term $(1/2)\boldsymbol{c}_{\tau}^\top \mathbb{V}\boldsymbol{c}_{\tau}$ for $\left\|\boldsymbol{c}_{\tau}\right\|$ greater than or equal to sufficiently large $\Lambda_{\tau}$.  Hence, (\ref{positiveprobability}) follows. 

Remains to show that (\ref{quadr_approx}) holds, uniformly in $\mathbf{c}_{\tau}$ in a compact set.  Observe that
\begin{eqnarray*}
\widehat{S}_{n}(\boldsymbol{c}_{\tau}/\sqrt{n}+(\gamma^*_{\tau},\boldsymbol{\eta}^*_{\tau}))&=& \frac{1}{2n}\boldsymbol{c}_{\tau}^{\top}\sum_{i=1}^{n}(1,\widehat{\mathbf{g}}(\mathbf{X}_{i}))(1,\widehat{\mathbf{g}}^{\top}(\mathbf{X}_{i}))^{\top} \boldsymbol{c}_{\tau}-\frac{1}{\sqrt{n}}\sum_{i=1}^{n} \widehat{Q}_{\tau}\{Y|\widehat{\mathbf{\Gamma}}^{\top}\widehat{\mathbf{g}}(\mathbf{X}_{i})\}(1,\widehat{\mathbf{g}}^{\top}(\mathbf{X}_{i}))^{\top}\boldsymbol{c}_{\tau}\\
&&-\sum_{i=1}^{n}\widehat{Q}_{\tau}\{Y|\widehat{\mathbf{\Gamma}}^{\top}\widehat{\mathbf{g}}(\mathbf{X}_{i})\}(1,\widehat{\mathbf{g}}^{\top}(\mathbf{X}_{i}))^{\top}(\gamma^*_{\tau},\boldsymbol{\eta}^*_{\tau})\\
&&+\frac{1}{2}(\gamma^*_{\tau},\boldsymbol{\eta}^{*\top}_{\tau})^{\top}\sum_{i=1}^{n}(1,\widehat{\mathbf{g}}(\mathbf{X}_{i}))(1,\widehat{\mathbf{g}}^{\top}(\mathbf{X}_{i}))^{\top}(\gamma^*_{\tau},\boldsymbol{\eta}^*_{\tau})\\
&=&\frac{1}{2}\boldsymbol{c}^{\top}_{\tau}\mathbb{V}_{n}\boldsymbol{c}_{\tau}+\mathbf{W}_{\tau,n}^{\top}\boldsymbol{c}_{\tau}+C_{\tau,n},
\end{eqnarray*} 
where $\mathbb{V}_{n}=n^{-1}\sum_{i=1}^{n}(1,\widehat{\mathbf{g}}(\mathbf{X}_{i}))(1,\widehat{\mathbf{g}}^{\top}(\mathbf{X}_{i}))^{\top}$, and $\mathbf{W}_{\tau,n}$ and $C_{\tau,n}$ are defined in (\ref{Wn}) and (\ref{Cn}), respectively.  It is easy to see that $\mathbb{V}_{n}=\mathbb{V}+o_{p}(1)$, and therefore, 
\begin{eqnarray*}
\widehat{S}_{n}(\boldsymbol{c}_{\tau}/\sqrt{n}+(\gamma^*_{\tau},\boldsymbol{\eta}^*_{\tau}))=\frac{1}{2}\boldsymbol{c}_{\tau}^{\top}\mathbb{V}\boldsymbol{c}_{\tau}+\mathbf{W}_{\tau,n}^{\top}\boldsymbol{c}_{\tau}+C_{\tau,n}+o_{p}(1).
\end{eqnarray*}
Provided that $\mathbf{W}_{\tau,n}$ is stochastically bounded, it follows from the convexity lemma (Pollard 1991) that the quadratic approximation to the convex function $\widehat{S}_{n}(\boldsymbol{c}_{\tau}/\sqrt{n}+(\gamma_{\tau}^*,\boldsymbol{\eta}^*_{\tau}))$ holds uniformly for $\boldsymbol{c}_{\tau}$ in a compact set.  Remains to prove that $\mathbf{W}_{\tau,n}$ is stochastically bounded.

Since $\mathbf{W}_{\tau,n}$ involves the quantity $\widehat{Q}_{\tau}\{Y|\widehat{\mathbf{\Gamma}}^{\top}\widehat{\mathbf{g}}(\mathbf{X}_{i})\}$, which is data dependent and not deterministic function, we define 
\begin{eqnarray*}
\mathbf{W}_{\tau,n}(\phi_{\tau})=-\frac{1}{\sqrt{n}}\sum_{i=1}^{n}\phi_{\tau}\{Y|\mathbf{\Gamma}^{\top}\mathbf{g}(\mathbf{X}_{i})\}(1,\widehat{\mathbf{g}}(\mathbf{X}_{i})),
\end{eqnarray*}
where $\phi_{\tau}: \mathbb{R}^{d_{2}+1} \rightarrow \mathbb{R}$ is a function in the class $\Phi_{\tau}$, whose value at $(y,\mathbf{\Gamma}^\top \mathbf{g}(\mathbf{x})) \in \mathbb{R}^{d_{2}+1}$ can be written as $\phi_{\tau}\{y|\mathbf{\Gamma}^{\top}\mathbf{g}(\mathbf{x})\}$, in the non-separable space $l^{\infty}(y,\mathbf{\Gamma}^{\top}\mathbf{g}(\mathbf{x}))=\{(y,\mathbf{\Gamma}^{\top}\mathbf{g}(\mathbf{x})): \mathbb{R}^{d_{2}+1} \rightarrow \mathbb{R}: \left\|\phi _{\tau}\right\|_{(y,\mathbf{\Gamma}^{\top}\mathbf{g}(\mathbf{x}))}:= \sup_{(y,\mathbf{\Gamma}^{\top}\mathbf{g}(\mathbf{x})) \in \mathbb{R}^{d_{2}+1}} |\phi_{\tau}(y|\mathbf{\Gamma}^{\top}\mathbf{g}(\mathbf{x}))|<\infty\}$, and satisfying $E|\phi _{\tau}\{Y|\mathbf{\Gamma}^{\top}\mathbf{g}(\mathbf{X})\}|^2 < \infty$ and $E \left\| \phi _{\tau}\{Y|\mathbf{\Gamma}^{\top}\mathbf{g}(\mathbf{X})\}^2 \mathbf{g}(\mathbf{X})\mathbf{g}^{\top}(\mathbf{X}) \right\| < \infty$.  Since $\Phi_{\tau}$ includes $Q_{\tau}\{Y|\mathbf{\Gamma}^{\top}\mathbf{g}(\mathbf{x})\}$, and, according to Lemma \ref{LemmaB1}, includes $\widehat{Q}_{\tau}\{Y|\widehat{\mathbf{\Gamma}}^{\top}\widehat{\mathbf{g}}(\mathbf{x})\}$ for $n$ large enough, almost surely, we will prove that $\mathbf{W}_{\tau,n}(\phi_{\tau})$ is stochastically bounded, \textit{uniformly} on $\phi_{\tau} \in \Phi_{\tau}$.

Observe that
\begin{eqnarray*}
&&\sup _{\phi_{\tau} \in \Phi_{\tau}} \left\| E \left\{\mathbf{W}_{\tau,n}(\phi_{\tau}) \mathbf{W}^{\top}_{\tau,n}(\phi_{\tau}) \right\} \right\| \\
 &\leq& \sup _{\phi_{\tau} \in \Phi_{\tau}} \frac{1}{n} \sum_{i=1}^{n} E \left[ \phi_{\tau}\{Y|\mathbf{\Gamma}^{\top}\mathbf{g}(\mathbf{X}_{i})\}^2 \left\| (1,\widehat{\mathbf{g}}(\mathbf{X}_{i}))(1,\widehat{\mathbf{g}}^{\top}(\mathbf{X}_{i}))^{\top} \right\| \right]\\
 &\leq& \sup _{\phi_{\tau} \in \Phi_{\tau}} \frac{1}{n} \sum_{i=1}^{n} E \left[ \phi_{\tau}\{Y|\mathbf{\Gamma}^{\top}\mathbf{g}(\mathbf{X}_{i})\}^2 \left\| (1,\widehat{\mathbf{g}}(\mathbf{X}_{i}))(1,\widehat{\mathbf{g}}^{\top}(\mathbf{X}_{i}))^{\top}-(1,\mathbf{g}(\mathbf{X}_{i}))(1,\mathbf{g}^{\top}(\mathbf{X}_{i}))^{\top} \right\| \right]\\
 &&+ \sup _{\phi_{\tau} \in \Phi_{\tau}} \frac{1}{n} \sum_{i=1}^{n} E \left[ \phi_{\tau}\{Y|\mathbf{\Gamma}^{\top}\mathbf{g}(\mathbf{X}_{i})\}^2 \left\| (1,\mathbf{g}(\mathbf{X}_{i}))(1,\mathbf{g}^{\top}(\mathbf{X}_{i}))^{\top} \right\| \right]\\
&=&O[E [ \phi_{\tau}\{Y|\mathbf{\Gamma}^{\top}\mathbf{g}(\mathbf{X})\}^2]]O_{p}(n^{-1/2})+ O\left[E \left[ \phi_{\tau}\{Y|\mathbf{\Gamma}^{\top}\mathbf{g}(\mathbf{X})\}^2\left\| (1,\mathbf{X})(1,\mathbf{X})^{\top} \right\| \right]\right]\\
&=&O(1),
\end{eqnarray*}
which follows from the properties of the class $\Phi_{\tau}$ defined above.  Bounded second moment implies that $\mathbf{W}_{\tau, n}(\phi_{\tau})$ is stochastically bounded.  Since
\begin{enumerate}
\item the result was proven uniformly on $\phi_{\tau}$,
\item the class $\Phi_{\tau}$ includes $\widehat{Q}_{\tau}\{Y|\widehat{\mathbf{\Gamma}}^{\top}\widehat{\mathbf{g}}(\mathbf{x})\}$ for $n$ large enough, almost surely, 
\end{enumerate}
the proof follows.

\subsection{Proof of Theorem \ref{thm1}} \label{AppendixB2}

Let $\widehat{\mathbf{W}}_{\tau}=(\widehat{\boldsymbol{\eta}}_{\tau,0}, \dots, \widehat{\boldsymbol{\eta}}_{\tau, p-1})$ be a $p \times p$ matrix, where $\widehat{\boldsymbol{\eta}}_{\tau,0}$ is the OLS slope estimate for the regression of $\widehat{Q}_{\tau}\{Y|\widehat{\mathbf{\Gamma}}^{\top}\widehat{\mathbf{g}}(\mathbf{X})\}$ on $\widehat{\mathbf{g}}(\mathbf{X})$ and $\widehat{\boldsymbol{\eta}}_{\tau,j}=E_{n}[\widehat{Q}_{\tau}\{Y|\widehat{\boldsymbol{\eta}}^{\top}_{\tau,j-1}\widehat{\mathbf{g}}(\mathbf{X})\}\widehat{\mathbf{g}}(\mathbf{X})]$, $j=1, \dots, p-1$.  Moreover, let $\mathbf{W}_{\tau}$ be the population level version of $\widehat{\mathbf{W}}_{\tau}$, that is, $\mathbf{W}_{\tau}=(\boldsymbol{\eta}_{\tau,0},\dots, \boldsymbol{\eta}_{\tau,p-1})$, where $\boldsymbol{\eta}_{\tau,0}$ is the OLS slope for the regression of $Q_{\tau}\{Y|\boldsymbol{\Gamma}^{\top}\mathbf{g}(\mathbf{X})\}$ on $\mathbf{g}(\mathbf{X})$, and $\boldsymbol{\eta}_{\tau,j}=E[Q_{\tau}\{Y|\boldsymbol{\eta}_{\tau,j-1}^{\top}\mathbf{g}(\mathbf{X})\}\mathbf{g}(\mathbf{X})]$.  It is easy to see that $\widehat{\mathbf{W}}_{\tau}$ converges to $\mathbf{W}_{\tau}$ at $\sqrt{n}$-rate.  This follows from Lemma \ref{LemmaB2} and the central limit theorem.  Then, for $\left\| \cdot \right\|$ the Frobenius norm,
\begin{eqnarray*}
\left\| \widehat{\mathbf{W}}_{\tau} \widehat{\mathbf{W}}_{\tau}^{\top} - \mathbf{W}_{\tau} \mathbf{W}_{\tau}^{\top} \right\| & \leq & \left\| \widehat{\mathbf{W}}_{\tau} \widehat{\mathbf{W}}_{\tau}^{\top} - \widehat{\mathbf{W}}_{\tau} \mathbf{W}_{\tau}^{\top}  \right\|\\
&& + \left\| \widehat{\mathbf{W}}_{\tau} \mathbf{W}_{\tau}^{\top} - \mathbf{W}_{\tau} \mathbf{W}_{\tau}^{\top} \right\| \\
& =& O_{p}(n^{-1/2}),
\end{eqnarray*}   
and the eigenvectors of $\widehat{\mathbf{W}}_{\tau} \widehat{\mathbf{W}}_{\tau}^{\top}$ converge to the corresponding eigenvectors of $\mathbf{W}_{\tau} \mathbf{W}_{\tau}^{\top}$.  Finally, according to Theorem 5 of Christou (2019), the subspace spanned by the $d_{Q_{\tau}\{Y|\mathbf{g}(\mathbf{X})\}}$ eigenvectors of $\mathbf{W}_{\tau}\mathbf{W}_{\tau}^{\top}$ falls into $\mathcal{S}_{Q_{\tau}\{Y|\mathbf{g}(\mathbf{X})\}}$ and the proof is complete.


\begin{thebibliography}{9}

\bibitem{Chaudhuri} Chaudhuri, P. (1991) Nonparametric estimates of regression quantiles and their local Bahadur representation. \emph{The Annals of Statistics} 19(2), 760--777. 

\bibitem{Christou b} Christou, E. (2019) Central Quantile Subspace.  Under revision paper.  

\bibitem{Christou and Akritas 2016} Christou, E. and Akritas, M.G. (2016) Single index quantile regression for heteroscedastic data. \emph{Journal of Multivariate Analysis} 150, 169--182.

\bibitem{Christou and Akritas 2018} Christou, E. and Akritas, M.G. (2018) Variable selection in heteroscedastic single index quantile regression.  \emph{Communication in Statistics - Theory and Methods} 47(24), 6019--6033.  

\bibitem{Guerre} Guerre, E. and Sabbah, C. (2012) Uniform bias study and Bahadur representation for local polynomial estimators of the conditional quantile function. \emph{Econometric Theory} 28(01), 87--129.

\bibitem{Koenker and Bassett} Koenker, R. and Bassett, G. (1978) Regression quantiles. \emph{Econometrica}, 46(1), 33--50. 

\bibitem{Kong et al.} Kong, E., Linton, O. and Xia, Y. (2010) Uniform Bahadur representation for local polynomial estimates of M-regression and its application to the additive model. \emph{Econometric Theory} 26(5), 1529--1564.   

\bibitem{Kong and Xia 2012} Kong, E. and Xia, Y. (2012) A single-index quantile regression model and its estimation.  \emph{Econometric Theory} 28(4), 730--768.

\bibitem{Kong} Kong, E. and Xia, Y. (2014) An adaptive composite quantile approach to dimension reduction.  \emph{The Annals of Statistics} 42(4), 1657--1688.

\bibitem{Li} Li, K.-C. (1991) Sliced inverse regression for dimension reduction. \emph{Journal of the American Statistical Association} 86(414), 316--327.

\bibitem{Artemiou} Li, B., Artemiou, A. and Li, L. (2011) Principal Support Vector Machines for linear and nonlinear sufficient dimension reduction. \emph{The Annals of Statistics} 39(6), 3182--3210.  
 
\bibitem{Li et al.} Li, B., Zhang, H. and Chiaromonte, F. (2005) Contour regression: a general approach to dimension reduction. \emph{Annals of Statistics} 33(4), 1580--1616.

\bibitem{Luo et al.} Luo, W., Li, B. and Yin, X. (2014) On efficient dimension reduction with respect to a statistical functional of interest. \emph{The Annals of Statistics} 42(1), 382--412.

\bibitem{Wang et al.} Wang, T., Guo, X. and Zhu, L. (2014) Transformed sufficient dimension reduction. \emph{Biometrika} 101(4), 815--829. 

\bibitem{Wu} Wu T.Z., K. Yu and Y. Yu, (2010) Single index quantile  regression. \emph{Journal  of Multivariate Analysis} 101, 1607--1621.   

\bibitem{Yu} Yu, K. and Jones, M. C. (1998) Local linear quantile regression. \emph{Journal of the American Statistical Association} 93(441), 228--238.

\end{thebibliography}
\end{document}